\documentclass[iop,revtex4]{emulateapj}

\newcommand*{\teff}{$T_{\rm eff}$}
\newcommand*{\logg}{$\log~g$}
\newcommand*{\feh}{[Fe/H]}

\newcommand*{\cfe}{[C/Fe]}
\newcommand*{\kms}{km s$^{-1}$}
\newcommand*{\zmax}{$Z_{\rm max}$}
\newcommand*{\rapo}{$r_{\rm apo}$}
\newcommand*{\rperi}{$r_{\rm peri}$}
\newcommand*{\vphi}{$V_{\rm \phi}$}

\newcommand*{\rsun}{$R_\odot$}

\usepackage{hyperref}
\usepackage{lscape}

\hypersetup{
    bookmarks=true,         
    unicode=false,          
    pdftoolbar=true,        
    pdfmenubar=true,        
    pdffitwindow=true,     
    pdfstartview={FitH},    
    pdftitle={},    
    pdfauthor={Beers},     
    pdfsubject={Astronomy},   
    pdfcreator={dvipdf},   
    pdfproducer={dvipdf}, 
    pdfkeywords={metal-poor stars},
    pdfnewwindow=true,      
    colorlinks=true,       
    linkcolor=red,          
    citecolor=blue,        
    filecolor=magenta,      
    urlcolor=cyan,           
    breaklinks=true,
    linktocpage
}

\begin{document}
\title{Bright Metal-Poor Stars from the Hamburg/ESO Survey.  II. \\ 
A Chemodynamical Analysis} 

\author{Timothy C. Beers, Vinicius M. Placco, Daniela Carollo}
\affil{Department of Physics and JINA Center for the Evolution of the Elements,\\
University of Notre Dame, Notre Dame, IN 46556, USA}
\email{tbeers@nd.edu,vplacco@nd.edu,dcarollo@nd.edu} 

\author{Silvia Rossi}
\affil{Instituto de Astronomia,  Geof\'{i}sica e Ci\^{e}ncias Atmosf\'{e}ricas, 
Departamento de Astronomia, Universidade de S\~{a}o Paulo,  
Rua do Mat\~{a}o  1226, 05508-900 S\~{a}o Paulo, Brazil}
\email{rossi@astro.iag.usp.br}

\author{Young Sun Lee}
\affil{Department of Astronomy \& Space Science, Chungnam National University,\\
Daejeon 34134, Republic of Korea}
\email{youngsun@cnu.ac.kr}

\author{Anna Frebel}
\affil{Massachussetts Institute of Technology and Kavli Institute 
for Astrophysics and Space Research, 77 Massachusetts Avenue, Cambridge, 
MA, 02139, USA}
\email{afrebel@mit.edu}

\author{John E. Norris}
\affil{Research School of Astronomy and Astrophysics, The
Australian National University,\\ Mount Stromlo Observatory, Cotter
Road, Weston, ACT 2611, Australia} 
\email{john.norris@anu.edu.au}

\author{Sarah Dietz}
\affil{Department of Physics and JINA Center for the Evolution of the Elements,\\
University of Notre Dame, Notre Dame, IN 46556, USA}
\email{sdietz@nd.edu}

\author{Thomas Masseron}
\affil{Institute of Astronomy, University of Cambridge,\\
Madingley Road, CB3 0HA, Cambridge, United Kingdom}
\email{tpm40@ast.cam.ac.uk}

\begin{abstract} 

We obtain estimates of stellar atmospheric parameters for a
previously published sample of 1777 relatively bright (9 $< B <$ 14)
metal-poor candidates from the Hamburg/ESO Survey. The original Frebel
et al. analysis of these stars was only able to derive estimates of
[Fe/H] and [C/Fe] for a subset of the sample, due to limitations in the
methodology then available. A new spectroscopic analysis pipeline has
been used to obtain estimates of \teff, \logg, \feh,
and [C/Fe] for almost the entire dataset. This sample is very local --
about 90\% of the stars are located within 0.5 kpc of the Sun. We
consider the chemodynamical properties of these stars in concert with a
similarly local sample of stars from a recent analysis of the Bidelman
\& MacConnell `weak-metal' candidates by Beers et al. We use this 
combined sample to identify possible members of the suggested halo stream of
stars by Helmi et al. and Chiba \& Beers, as well as stars that may
be associated with stripped debris from the putative parent dwarf of
the globular cluster Omega Centauri, suggested to exist by previous
authors. We identify a clear increase in the cumulative frequency of
carbon-enhanced metal-poor (CEMP) stars with declining metallicity, as
well as an increase in the fraction of CEMP stars with distance from the
Galactic plane, consistent with previous results. We also identify a
relatively large number of CEMP stars with kinematics consistent with
the metal-weak thick-disk population, with possible implications for its
origin. 

\end{abstract}

\keywords{stars: abundances -- stars: Population II -- stars: stellar dynamics
-- stars: weak-line -- Galaxy: kinematics and dynamics -- Galaxy: stellar
content -- Galaxy: structure}

\section{Introduction}

There have been numerous recent studies of the disk system of the Milky
Way, primarily based on data from the Sloan Digital Sky Survey
\citep[SDSS;][]{york2000}, in particular the SEGUE \citep{yanny2009}
and APOGEE \citep{majewski2015} sub-surveys, as well as from the Radial
Velocity Experiment \citep[RAVE; ][]{steinmetz2006,kordopatis2013} and
the Gaia-ESO survey \citep{gilmore2012,guiglion2015}. \citet{beers2014}
and \citet{guiglion2015} summarize the pertinent papers, to which the 
interested reader is referred. Most of these papers model the Galactic
disk system in terms of a superposition of a thin disk, a thick disk,
and (in some cases) a metal-weak thick disk (MWTD). The series of papers
from Bovy and collaborators, culminating with \citet{bovy2015}, has
taken a different approach. These authors use abundance information
([Fe/H] and [$\alpha$/Fe]) for large samples of red-clump stars measured
with APOGEE to model the radial and vertical structure of the disk in
terms of mono-abundance populations (MAPs), and demonstrate that this
technique captures the relevant observations without invoking a
separation of stellar populations. Because the MAPs are based on
red-clump stars, they do not include any stars with [Fe/H] $< -1.0$, and
so are not suitable for exploring issues relating to the MWTD, which
\citet{beers2014} have argued is a potentially separate component of the
disk system that has yet to be explored in detail. The recent paper by
\citet{kawata2016} emphasizes that the jury is still out concerning the
separability of the thin disk and thick disk -- these authors argue that
the scheme of chemically dividing the disk system on the basis of the
[$\alpha$/Fe] ratio, pioneered by \citet{lee2011a,lee2011b}, is, for now, the
most practical approach. 

One of the first large spectroscopic samples of stars in the disk system
was originally reported on by Frebel et al. (2006, hereafter Paper~I).
These stars, selected from partially saturated objective-prism spectra
from the Hamburg/ESO survey \citep[HES;][]{wisotzki2000,christlieb2003}
with 9 $< B <$ 14, formed the basis of an early effort to identify
bright metal-poor halo stars in the Galaxy. Due to flaws in the
selection procedure, the great majority of these stars turned out to
have metallicities more typical of the disk system rather than the halo.
Even so, the star HE~1327-2326, which was first identified in this
effort, turned out to have one of the lowest iron abundance known
(HE~1327-2326, with [Fe/H] = $-$5.45; \citealt{frebel2005, aoki2006}),
only recently surpassed by SMSS~J031300.36$-$670839.31, with [Fe/H] $<
-7.8$ \citep{keller2014,bessell2015}. This paper was also the first to
suggest an increase in the fraction of carbon-enhanced metal-poor (CEMP;
\citealt{beers2005}) stars with distance from the Galactic plane, which
was later confirmed with much larger samples of stars from SDSS
\citep{carollo2012}.

A substantial fraction of very low metallicity stars in the halo of the
Milky Way have been found to be CEMP stars. \cite{beers2005} originally
divided such stars into several sub-classes, depending on the nature of
their neutron-capture element abundance ratios -- CEMP-$s$, CEMP-$r$,
CEMP-$r/s$, and CEMP-no\footnote{CEMP-$s$ : [C/Fe]$>$+1.0, [Ba/Fe]$>$
+1.0, and [Ba/Eu]$>$+0.5\newline CEMP-$r$ : [C/Fe] $>$ +1.0 and
[Eu/Fe]$>$ +1.0\newline CEMP-$r/s$ : [C/Fe] $>$ +1.0 and 0.0 $<$ [Ba/Eu] $<$+0.5 
\newline CEMP-no : [C/Fe] $>$ +1.0 and [Ba/Fe] $<$ 0.0}. As
discussed by these authors, and many since, the observed differences in
the chemical signatures of the sub-classes of CEMP stars are thought to
arise due to differences in the astrophysical sites responsible for
the nucleosynthesis products they now incorporate in their
atmospheres, including elements produced by the very first generations 
of stars.

At the time Paper~I was published, the authors could only obtain
estimates of [Fe/H] and [C/Fe] from their spectra for
those stars with [Fe/H] $< -1.0$, due to nascent saturation of the 
\ion{Ca}{2}\,K line. This limitation precluded a comprehensive
investigation of the disk system including stars over the full range
of expected metallicities. Over the course of the past decade, we have
developed and tested new spectroscopic tools (primarily for application
to SDSS stellar spectra -- the SEGUE Stellar Parameter Pipeline; SSPP)
that are useful for the analysis of stars over wide ranges of [Fe/H]
(see, e.g., \citealt{lee2008a}). In this paper, we employ a modification
of the SSPP that can be used for spectra of similar resolving power, and
with input broadband $V, B-V$ and/or $J, J-K$ photometry, to obtain
estimates of the stellar atmospheric parameters \teff, \logg, and
[Fe/H], as well as [C/Fe] abundance ratios, for most of the stars in the
Paper~I sample. Similarly determined quantities from the local sample of 
`metal-weak' candidates from \citet{bidelman1973}, reported on recently 
by \citet{beers2014}, is analyzed in concert with the Paper~I sample. 

This information (in combination with well-determined radial velocities
and with available accurate proper motions) is employed to carry out a
detailed examination of the kinematics of the combined sample, identify
stars that are possible members of the halo stream/trail of stars by
\citet{helmi1999} and \citet{chiba2000}, as well as stars that may be 
associated with stripped debris from the putative parent dwarf of the
globular cluster Omega Centauri ($\omega$ Cen), suggested to exist by
\citet{dinescu2002}, \citet{klement2009}, and \citet{majewski2012}.
We identify a clear increase in the cumulative frequency of CEMP stars
as a function of declining metallicity, as well as an increase in the
fraction of CEMP stars with distance from the Galactic plane, as
quantified by the maximum distances reached during the course of their
orbits, \zmax, both consistent with previous results. We also
identify a number of CEMP stars that are apparently associated with
the MWTD, with implications for its origin. Finally, we make use of the
Yoon-Beers diagram of $A$(C) vs. [Fe/H] (Figure 1 of \citealt{yoon2016})
to sub-classify the relative small number of CEMP stars in the combined
sample with available kinematic information (36 stars) into likely
CEMP-$s$ and CEMP-no stars, and show that their distributions of \zmax\
differ, in the sense that the CEMP-$s$ stars appear to be preferentially
associated with the inner-halo population, while the CEMP-no stars are
more likely to be associated with the outer-halo population, similar to
the previous claim of \citet{carollo2014}.

\section{Sample Stars and Adopted Photometry}
\label{observations}

Paper~I describes the original motivations and selection of the
bright candidate metal-poor stars from the HES, to which
the interested reader is referred for details. Unfortunately, the
original candidate selection was confounded by the (known) saturation
effects on the derived estimates of approximate $B-V$ to such a degree
that numerous stars were included that later turned out to be more
metal-rich than hoped for. In spite of this limitation, more than a
hundred relatively bright very metal-poor (VMP; [Fe/H] $< -2.0$) stars
were identified during the follow-up spectroscopy, which formed the
basis for much of the analysis carried out in Paper~I. 

In the present paper, we re-analyze medium-resolution ($R \sim 2000$)
spectroscopy of the sample of stars from Paper~I (see Table 6 of Paper~I
for the telescope/spectrographs employed), using the n-SSPP
spectroscopic pipeline described below. This new effort, which also
incorporates a large amount of newly available broadband $V, B-V$
photometry for the sample stars, enables determinations of stellar
atmospheric parameters for the great majority of the Paper~I sample,
including stars with metallicities up to Solar and beyond (which was not
previously possible), as well as refined estimates of [C/Fe] abundance
ratios for most of the stars in this sample.

\subsection{Broadband Photometry and Reddening Estimation}

Broadband $V$ magnitudes and $B-V$ colors for the majority of our
program objects were obtained from the APASS database
\citep{henden2015}, supplemented by photometry from a number of 
sources as described in Paper~I (primarily stars that were re-discoveries
of metal-poor candidates from the HK survey; Beers et al. 1985, 1992).
For stars that were of particular interest, i.e., those found in Paper~I
to have [Fe/H] $< -2.0$ or that exhibited enhanced carbon, we
also make use of photometry reported by \citet{beers2007}. In a number of
cases, we have also used photometry from the SIMBAD database. For stars
with photometry available from multiple sources, we either used the data
judged to be superior, or else averaged data expected to be of similar
precision. Near-IR $JHK$ photometry from the 2MASS catalog
\citep{skrutskie2006} is available for all but a few stars in our sample.    

Column (1) of Table~\ref{tab1} lists the star names, while columns (2)
and (3) list other common names for the star and the HK Survey star
name, respectively. The full set of coordinates for our program stars
are provided in Paper~I. Columns (4) and (5) list the Galactic longitude
and latitudes for our program stars. The adopted $V$ magnitude and $B-V$
colors are provided in columns (6) and (7). The 2MASS $J$ magnitude and
$J-K$ colors (only including stars without flags indicating potential
problems in the listed values) are listed in columns (8) and (9). 

In order to obtain absorption- and reddening-corrected estimates of the
magnitudes and colors, respectively, we initially adopted the
\citet{schlegel1998} estimates of reddening listed in column (10) of
Table~\ref{tab1}. We have applied corrections to these estimates for
stars with reddening greater than $E(B-V)_S$ = 0.10, as described by
\citet{beers2000}. The corrected reddening estimates, $E(B-V)_A$, are
listed in column (11).

\section{Radial Velocities, Line Indices, Atmospheric Parameters, Abundance
Ratios, Distances, and Proper Motions}
\label{abundances}

\subsection{Measurement of Radial Velocities and Line Indices}

Radial velocities were (re-)measured for our program stars using the
line-by-line and cross-correlation techniques described in detail by
\citet{beers1999} and references therein.  In the process of carrying
out this exercise, we found that many of the new measurements differ (in
some cases by large amounts) from those originally reported in Paper~I,
which apparently suffered from transcription difficulties during file exchanges
between the authors. The current measurements supersede
those values. The spectral resolution of our data is similar to that
obtained for the majority of the HK Survey follow-up, thus we expect
that the measured radial velocities should be precise to the same level
(or better, given the higher signal-to-noise of our present spectra), on
the order of 7-10 km~s$^{-1}$ (one sigma). Column (1) of
Table~\ref{tab2} is the star name, while column (2) lists notes on the
nature of a small number of stars that deviate from the majority (e.g.,
hot stars, sdB and WD stars, known variable stars, stars with emission
lines, extremely late-type stars, etc.), which preclude their use in
later analysis. We conservatively estimate that our medium-resolution
velocities, RV$_M$, listed in column (3) of Table~\ref{tab2}, are
precise to 10~km~s$^{-1}$ (as validated below).

Roughly one-third of our program objects (614 stars) have had radial
velocities determined from the RAVE survey, based on moderate-resolution
($R \sim 7500$) spectroscopy in the region of the Ca triplet from Data
Release 4 \citep{kordopatis2013}. The RAVE velocities should be more
precise than those we obtained from our lower-resolution spectra
(\citealt{kordopatis2013} demonstrate that the majority of the RAVE
radial velocities are precise to better than 2 \kms, with a tail going
out to $\sim$5 \kms, and have small zero-point offsets relative to external
catalogs). For our purpose we conservatively assume a precision of 5
\kms\ for the RAVE velocities (validated below). We adopt the RAVE
velocities for our subsequent analysis, except in cases where flags were
raised in the DR4 database indicating potential problems (including
possible binary membership). The available RAVE velocities are listed as
RV$_R$ in column (4) of Table~\ref{tab2}. In order to weed out stars
with inaccurate RAVE velocities, we have indicated stars with flags
suggesting potential problems with parentheses around them. We still
adopt these radial velocities for our analysis if they are within 20
km~s$^{-1}$ of the listed RV$_M$ value. If the RAVE velocities differ by
more than this amount, and had flags raised, we assume that the RV$_M$
estimates are superior. We indicate such cases by brackets around the
listed RV$_R$ values. In some instances there are no flags raised, but the
RAVE radial velocities differ from our medium-resolution results by more
than 20 km~s$^{-1}$; in such cases we assume the RAVE estimates are
superior, and adopt them.  

There are 148 stars in our sample (mostly VMP stars, or stars of interest for
other reasons, such as carbon enhancement) for which radial velocities
based on high-resolution spectroscopy are available, either in the
published literature or based on more recent unpublished observations
we are aware of. These are listed as RV$_H$ in column (5) of
Table~\ref{tab2}. We adopt these measurements (when available), with
assumed errors of 2 km~s$^{-1}$, even in the few cases where they
disagree by more than 20 km~s$^{-1}$ with either the RV$_M$ or RV$_R$
velocities.  Unrecognized binarity
may be responsible for a number of these discrepancies.

\begin{figure*}[!ht]
\begin{center}
\includegraphics[scale=0.545]{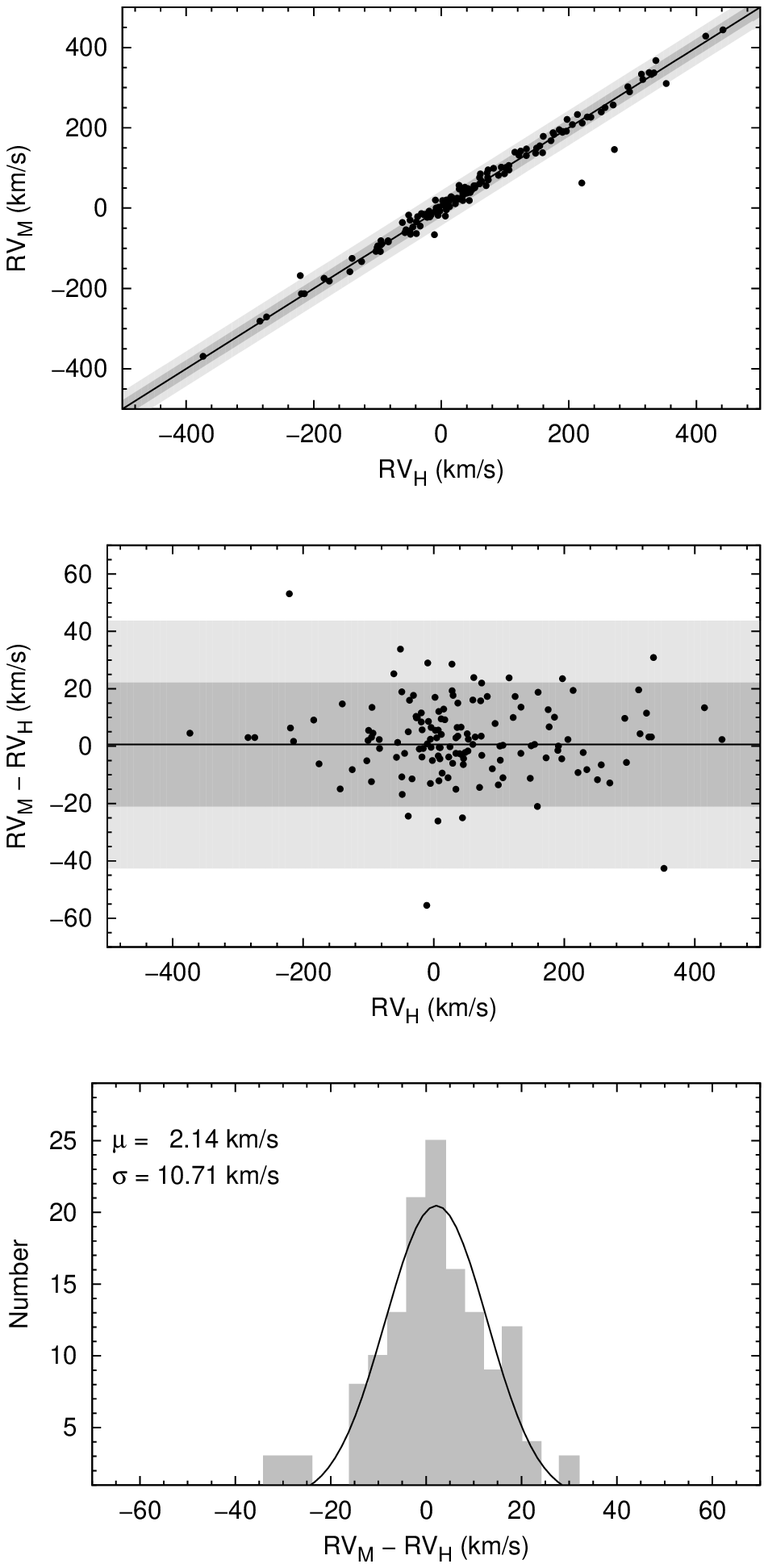}
\includegraphics[scale=0.545]{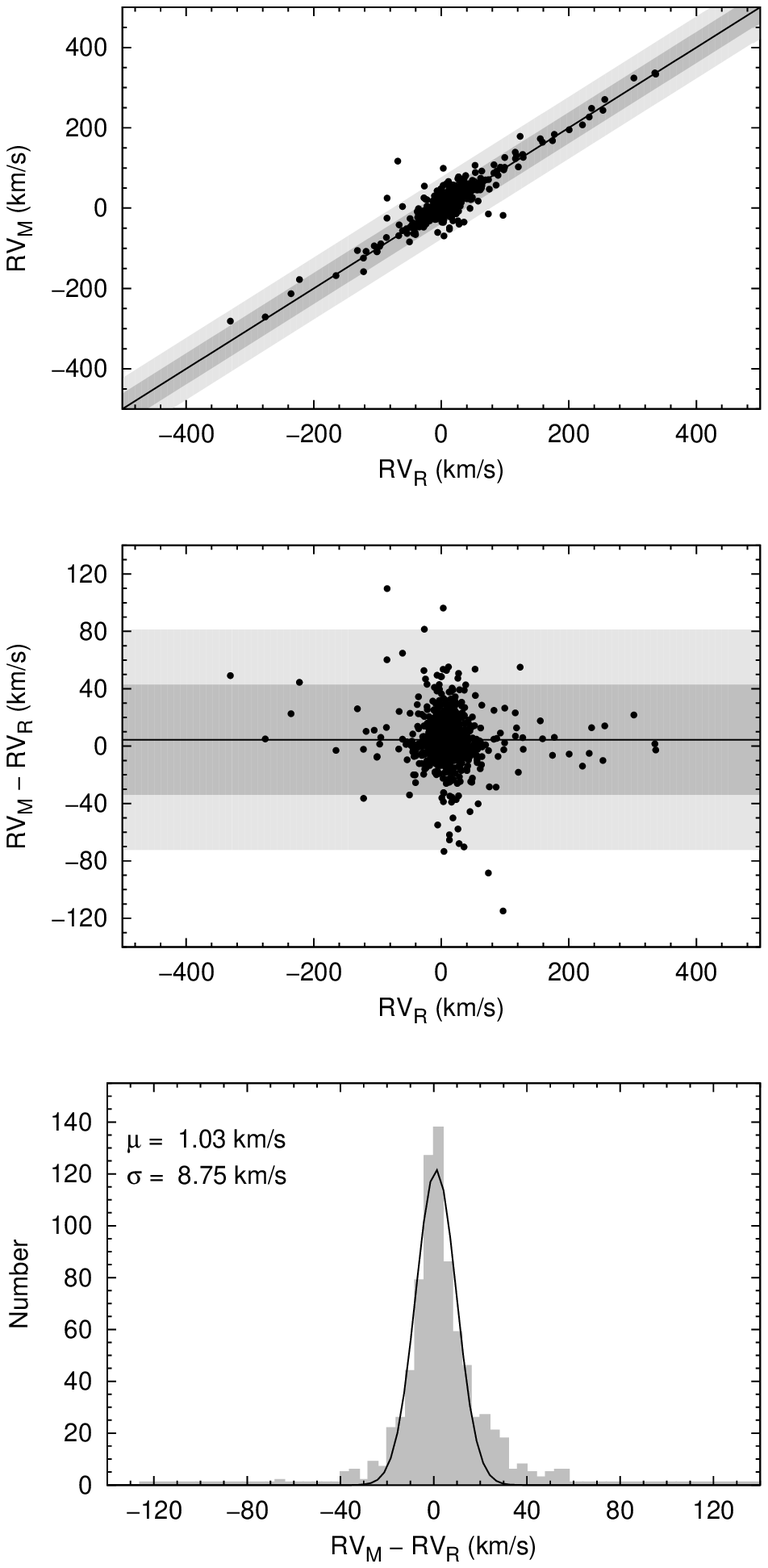}
\includegraphics[scale=0.545]{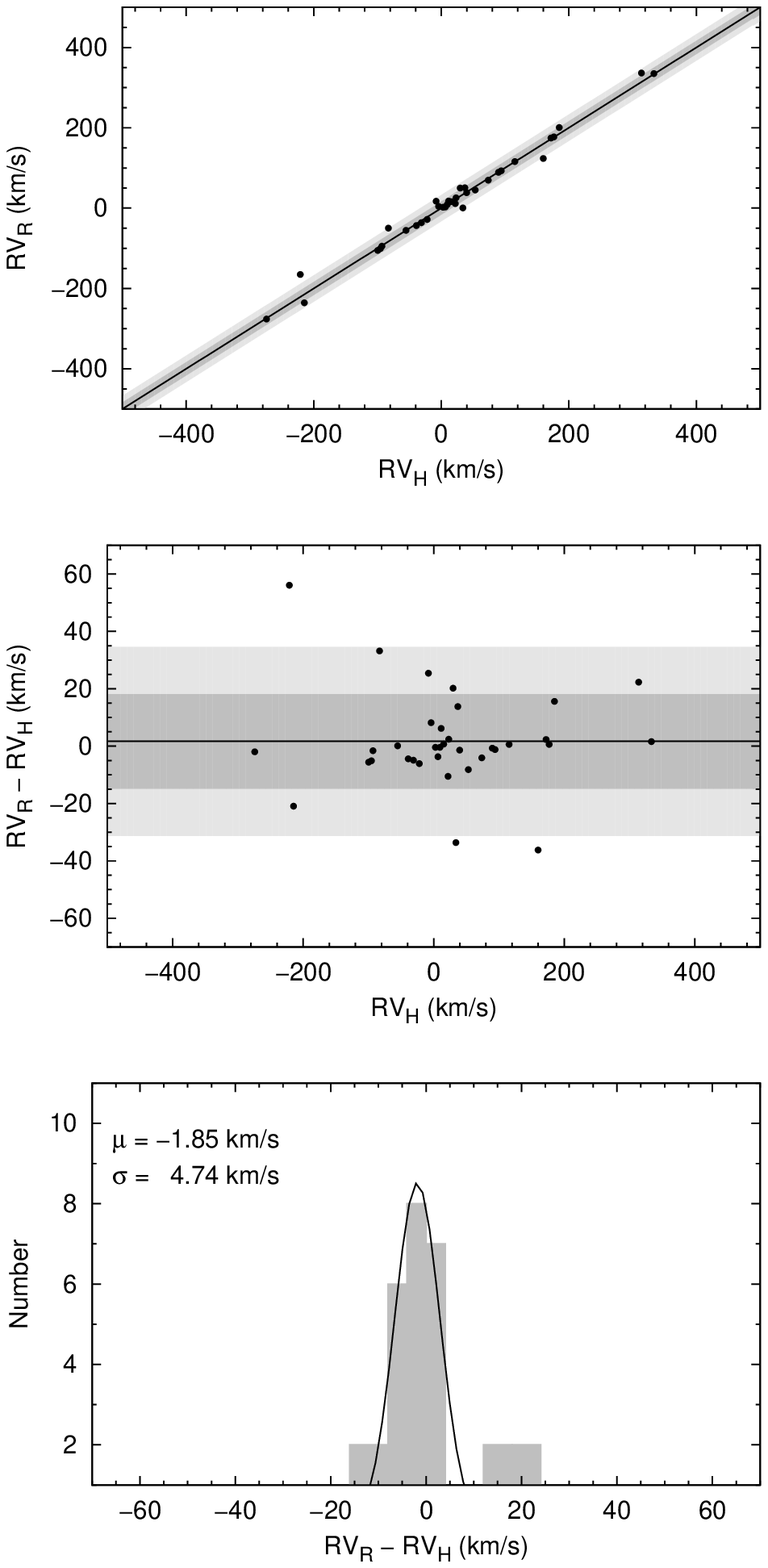}
\caption{Radial velocity comparison for the program stars (RV$_M$), RAVE
(RV$_R$), and high-resolution (RV$_H$). Upper panels: 
Comparison between the radial velocities. The solid line is the
one-to-one line, and the shaded areas represent the 1$\sigma$ and
2-$\sigma$ intervals around this line (where $\sigma$ represents the
scatter in the residuals shown in the lower panels. Middle panels:
Residuals between each pair of measurements. The horizontal solid line
is the average of the residuals, while the darker and lighter shaded
areas represent the 1-$\sigma$ and 2-$\sigma$ regions, respectively.
Lower panels: Histogram of the residuals in the radial-velocity
determinations. The values of the mean offset and scatter are the
parameters from the Gaussian fit shown.}
\label{rv}
\end{center}
\end{figure*}

Figure~\ref{rv} (left-hand column) compares the medium-resolution
velocities, RV$_M$, with the high-resolution radial velocities, RV$_H$,
while the middle column of panels compares the RV$_M$ velocities with
the moderate-resolution RAVE velocities, RV$_R$ (excluding the rejected
cases). The right-hand column of panels compares results of the RAVE
velocities with the high-resolution velocities. As can be appreciated
from inspection of this figure, there is generally very good agreement
between the different sources of radial velocity. The middle row of
panels shows the residuals in radial velocity for each comparison, with
dark gray and light gray regions indicating the 1-$\sigma$ and
2-$\sigma$ ranges, respectively. Maximum-likelihood fits to the
residuals in radial velocity for each comparison are shown in the lower
panels of each column. The RV$_M$ vs. RV$_H$ residuals exhibit a scatter
of 10.7 \kms\, and a small zero-point offset; the RV$_M$ vs. RV$_R$
residuals exhibit a scatter of 8.8 \kms\ and a similarly small offset.
The RV$_R$ vs. RV$_H$ residuals exhibit a scatter of 4.7 \kms\ and a
small offset. Assuming our adopted estimate of the 2 \kms\ precision for
the high-resolution radial velocities, our results indicate that the
RAVE radial velocities are precise to 4.2 \kms\ (note that this
comparison emphasizes metal-poor stars for which the RAVE velocities are
expected to be somewhat less precise than for more metal-rich stars).
Adopting this value for the scatter in the RAVE velocities, we estimate
that the medium-resolution velocities have a precision of 7.7 \kms. When
compared to the high-resolution velocities, we estimate that the
medium-resolution velocities have a precision of 10.5 \kms. These values
justify our adopted estimates for the kinematic analysis carried out
below -- $\sigma_{\rm{RV}_R}$ = 5 \kms\ and $\sigma_{\rm{RV}_M} = 10$
\kms, for the RAVE and medium-resolution velocities, respectively. 

For each star, the measured (geocentric) radial velocities are used to
place a set of fixed bands for the derivation of line-strength indices,
which are pseudo-equivalent widths of prominent spectral features. We
employ a subset of the bands listed in Table 1 of
\citet{beers1999}.\footnote{A complete discussion of the choice of bands,
the ``band-switching'' scheme, and the Balmer line index, HP2, which
measures the strength of the H$\delta$ lines, is provided in this
reference as well.} Although we do not make use of them in the present
analysis, others may choose to, so we list line indices for prominent
spectral features in each of our program stars in columns (6)-(8) of
Table~\ref{tab2}. A number of our stars have had more than one spectrum
obtained during the course of our follow-up observations. From a
comparison of the stars with repeated measurements, we estimate that
errors in the line indices on the order of 0.1 {\AA} are achieved. 
Note that our line indices are identical to those reported in Paper~I.

\clearpage

\subsection {Stellar Atmospheric Parameter Estimates and Abundance Ratios}

In a series of papers, \citet{lee2008a}, \citet{lee2008b},
\citet{allendeprieto2008}, \citet{lee2011a}, and \citet{smolinski2011}
describe the development, testing, and validation of the SSPP software,
which has been used to determine atmospheric parameter estimates for
over 500,000 stars from the SDSS and its extensions. Although the
spectra of our program stars do not reach as far red as SDSS spectra
(and hence we cannot use as many of the independent methods as the SSPP
enables), they are of a similar resolving power. We have thus modified
the SSPP to accept input from our program spectra, which span spectral
ranges of 3600-4400\,{\AA}, 3600-4800\,{\AA}, or 3600-5250\,{\AA},
depending on the telescope/spectrograph that was used to acquire them.
We have also implemented the use of input $V, B-V$ (and/or 2MASS $J,
J-K$) photometric information, rather then requiring SDSS $ugriz$
inputs. This new approach, known as the n-SSPP (for ``non-SEGUE''
Stellar Parameter Pipeline), makes use of a subset of previously
calibrated methods from the SSPP (those that apply to the available
wavelength range of the input spectra) to obtain estimates of the
fundamental stellar parameters \teff, \logg, and \feh. For spectra that
extend sufficiently red-ward to include the CH $G$-band at $\sim 4300$\,
{\AA}, and/or the \ion{Mg}{1} feature at $\sim 5175$\,{\AA}, the n-SSPP
can obtain estimates of [C/Fe] and [$\alpha$/Fe]\footnote{This notation
is usually defined by an average of [Mg/Fe], [Si/Fe], [Ca/Fe], and
[Ti/Fe].} as well (if spectra are of sufficiently high signal-to-noise,
S/N). The n-SSPP has already been applied by \citet{beers2014} to
medium-resolution spectra of stars from the sample of
\citet{bidelman1973} stars studied by \citet{norris1985}. The interested
reader should consult this paper for additional information on the
operation of the n-SSPP.

We apply the n-SSPP to the sample of 1777 stars from Paper~I.
Unfortunately, the S/N ratios for the spectra of our program stars that
extend to wavelength regions that include the \ion{Mg}{1} feature are
not generally high enough to enable confident estimation of this
abundance ratio (\citealt{lee2011a} recommend S/N $>$ 20 or 25, the
latter applying to stars with [Fe/H] $< -1.4$), hence we do not report
[$\alpha$/Fe] for our program stars.

The n-SSPP estimates of stellar atmospheric parameters are listed in
columns (9)-(11) of Table~\ref{tab2} as Teff$_S$, logg$_S$, and
[Fe/H]$_S$, respectively. Although, according to the tests described by
Lee et al. (2008a,b) and \citet{allendeprieto2008}, the external
accuracy of the SSPP parameter estimates are expected to be on the order
of 150~K, 0.30 dex, and 0.25 dex for \teff, \logg, and [Fe/H],
respectively, these are based on the availability of SDSS $ugriz$ and
the full spectral coverage associated with SDSS spectroscopy, neither of
which apply to the present data. We provide an independent test of our
expected parameter errors below.  

\citet{lee2013} describes the procedures adopted to estimate \cfe{} 
for SDSS/SEGUE spectra, based on spectral matching against a dense grid
of synthetic spectra; these techniques, with different input photometric
information, also apply to the n-SSPP. We have recently expanded the
carbon grid to reach as low as [C/Fe] $= -1.5$, rather than the limit of
[C/Fe] $= -0.5$ employed by \citet{lee2013}. According to Lee et al.,
the precision of the \cfe{} estimates is better than $0.35$~dex for the
parameter space and S/N ratios explored by SDSS/SEGUE spectra. We expect
improved results for application of the n-SSPP to our program spectra,
based on their generally higher signal-to-noise (which typically exceed
S/N $\sim 50$ in the region of the CH $G$-band). We note that
\citet{beers2014} concluded that the n-SSPP determination of [C/Fe] for
similar S/N spectra as our current program achieved a precision (based
on empirical comparisons with high-resolution spectroscopic analyses) of
$\sim 0.20$~dex. 

Table~\ref{tab3} lists the medium-resolution estimates of the [C/Fe]
abundance ratios (``carbonicity'') for our program stars in column (3),
indicated as [C/Fe]$_S$. For convenience, we have also listed the n-SSPP
estimate of [Fe/H]$_S$ in column (2). Column (4) indicates whether the
listed measurement is considered a detection, DETECT = ``D,'' lower
limit ``L,'' upper limit ``U,'' or a non-detection, ``X,'' indicating
that the star is either too hot (or cool) for carbon to be measured from
the CH $G$-band, or that the star does not have a reference metallicity
determination. Column (5) provides the correlation coefficient, CC,
obtained between the observed spectrum and the best-matching [C/Fe] from
the model grids, and column (6) lists the equivalent width of the CH
$G$-band, EQW. For an acceptable measurement of this ratio, we demand
DETECT = ``D,'' CC $\ge 0.7$, and EQW $\ge 1.2$. The latter restriction
insures that stars with very weak carbon features are not spuriously
assigned values by the grid search procedure. See \citet{lee2013} for
further discussion of these quantities. Stars for which either CC or EQW
does not meet the minimum value are indicated by a colon attached to the
DETECT parameter in column (4). There are 1491 stars listed in
Table~\ref{tab3} for which acceptable measurements of [C/Fe] are
obtained -- 1422 are listed as detections, 58 as upper limits, and 11 as
lower limits. 


\subsection{Comparison to Moderate- and High-Resolution Spectroscopic Analyses}

There are external measurements of stellar atmospheric-parameter
estimates for 707 stars in our sample from the RAVE DR4 catalog
\citep{kordopatis2013}, and another 104 stars for which atmospheric
parameter estimates based on high-resolution analyses are available from
a variety of sources, including the SAGA database \citep{suda2008,
suda2011,yamada2013} and \citet{frebel2010}, as well as the references
listed in the PASTEL catalogue \citep{soubiran2010}\footnote{We are
aware that an updated version of this catalogue is now available,
\citet{soubiran2016}, but it was published after we completed the bulk
of our analysis, and hence not used for this exercise.}, supplemented by
other determinations that have appeared in more recent studies, or
unpublished results from co-authors of this paper. We have adopted
either the parameter estimates we judged to be superior, or, in some
cases, took a straight average of the available estimates.

The external parameter estimates from RAVE are listed in columns 
(12)-(14) of Table~\ref{tab2} as Teff$_R$, log g$_R$, and [Fe/H]$_R$,
respectively. The high-resolution estimates are listed in columns 
(15)-(17) of this table as Teff$_H$, log g$_H$, and [Fe/H]$_H$,
respectively.  

Before carrying out comparisons with our own estimates, based on
medium-resolution spectra, we first check for external parameter
estimates that grossly differ from the estimates determined by the
n-SSPP. In order for the external estimates to be considered
commensurate with the n-SSPP estimates, reasonable agreement for
effective temperature, \teff, is required, at a minimum. To implement
this pre-filter, we demand that the estimated effective temperatures
from the external comparisons are within 500~K of the n-SSPP estimates,
and (in the case of RAVE) that there be no other indication of potential
problems, such as flags raised in the RAVE DR4 catalog listing. This
results in a total of 80 stars with RAVE estimates being marked as
suspect, indicated in Table 2 with brackets around the individual
parameter estimates. Only 9 stars with available high-resolution
spectroscopic parameter estimates are suspect, by this criterion, and
these are marked with parentheses around the individual parameter
estimates in the table. 

\begin{figure*}[!ht]
\epsscale{1.00}
\plotone{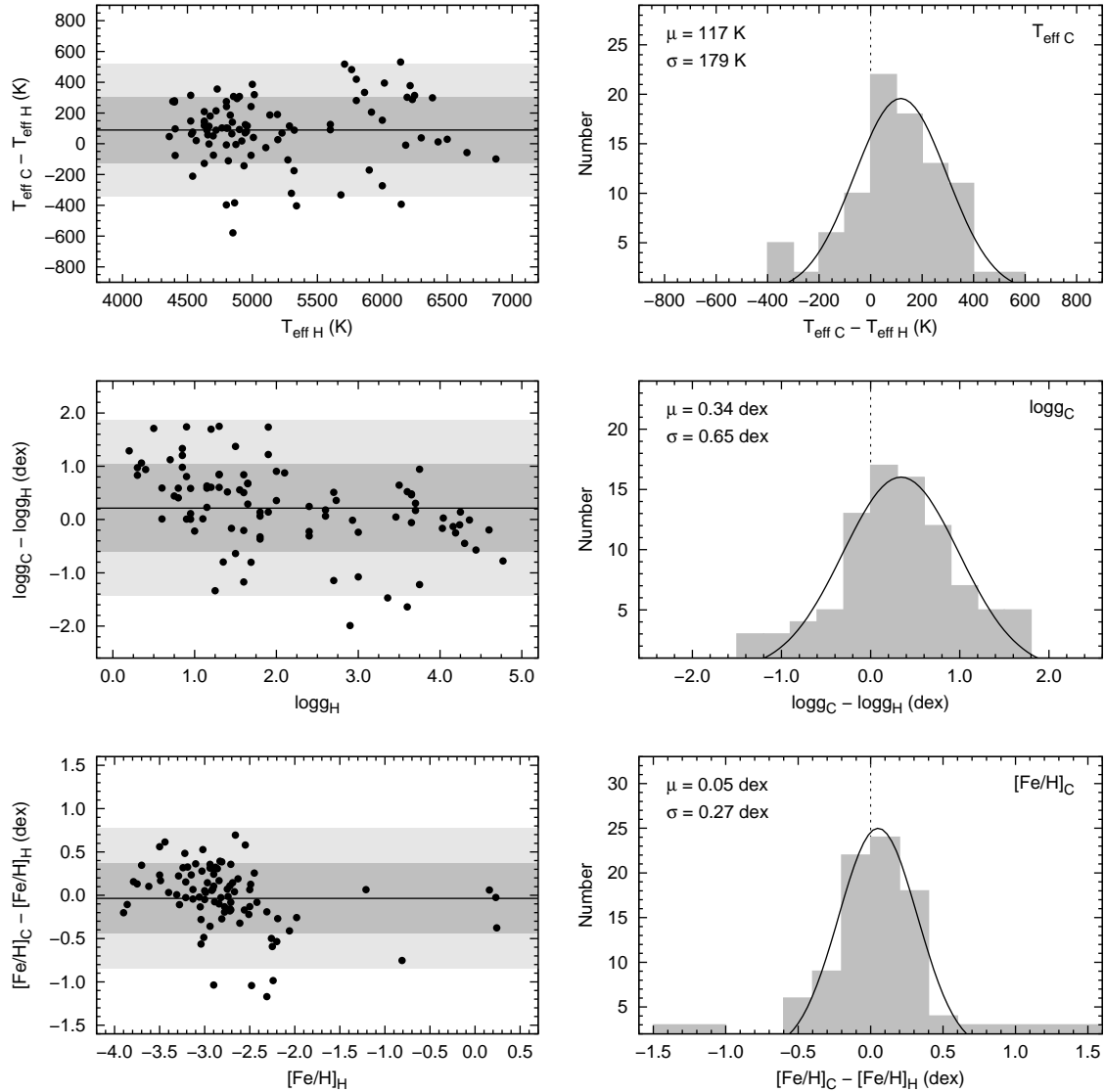}
\caption{Left panels: Differences between the (corrected) atmospheric parameters 
determined by the n-SSPP, Teff$_C$, logg$_C$, and [Fe/H]$_C$, and the values
from analyses of high-resolution spectroscopy, Teff$_H$, logg$_H$, and
[Fe/H]$_H$,  reported in the literature, as a function
of the high-resolution spectroscopic values.  Filled symbols refer
to the program stars. The horizontal solid line is the average
of the residuals, while the darker and lighter shaded areas represent
the 1-$\sigma$ and 2-$\sigma$ regions, respectively.
Right panels: Histograms of the residuals between the
corrected n-SSPP and high-resolution parameters shown in the left
panels. Each panel also lists the average offset and scatter
determined from a Gaussian fit.}
\label{atmMH}
\end{figure*}

\citet{beers2014} presented a similar analysis for the sample of
302 metal-poor candidates from \citet{bidelman1973} studied by
\citet{norris1985}, for which roughly one-third of the sample had
external estimates of stellar atmospheric parameters based on
high-resolution spectroscopic analyses. Beers et al. used the sample of
stars in common to derive empirical corrections to the n-SSPP parameter
estimates, which they applied in order to place these estimates on a
scale commensurate with the high-resolution work. For convenience of the
reader, these corrections are listed below:

\begin{eqnarray}
{\rm [Fe/H]}_C     = {\rm [Fe/H]}_S - ( -0.232 \cdot {\rm [Fe/H]}_S - 0.428 )  \\
{\rm Teff}_C   = {\rm Teff}_S -  ( -0.1758 \cdot {\rm Teff}_S + 1062)  \\
{\rm logg}_C       = {\rm logg}_S -  ( -0.237 \cdot {\rm logg}_S + 0.523 ) 
\label{eqcor}
\end{eqnarray}

\noindent The corrected n-SSPP estimates (Teff$_C$, logg$_C$, and [Fe/H]$_C$)
for our program stars are listed in columns (18)-(20) of Table~\ref{tab2}.
Column (21) of this table lists our adopted type classifications,
obtained as described below.  

Figure~\ref{atmMH} illustrates comparisons of Teff$_C$, logg$_C$, and
[Fe/H]$_C$ for our program stars with the adopted high-resolution
results. Note that, with the exception of a few individual stars lying
outside the 2-$\sigma$ bands shown in the left-hand panels, the
agreement is quite satisfactory. Maximum-likelihood fits to the
distributions of residuals between these various estimates are shown in
the right-hand panels. Both the mean offsets ($\Delta$~Teff$_C$ $=
117$~K, $\Delta$~logg$_C$ $= 0.34$~dex, $\Delta$~[Fe/H]$_C$ $=
0.05$~dex) and the scatter in the estimates ($\sigma$~Teff$_C$ $=
179$~K, $\sigma$~logg$_C$ $= 0.65$~dex, $\sigma$~[Fe/H]$_C$ $=
0.27$~dex) are reasonably small. Taking into account the expected errors
in the high-resolution estimates of these parameters (125~K, 0.4 dex,
and 0.2 dex, respectively), we conclude that the external precisions of
the n-SSPP estimates of Teff$_C$, logg$_C$, and [Fe/H]$_C$ are on the
order of 125~K, 0.5~dex, and 0.2~dex, respectively. 

\begin{figure*}[!ht]
\epsscale{1.15}
\plotone{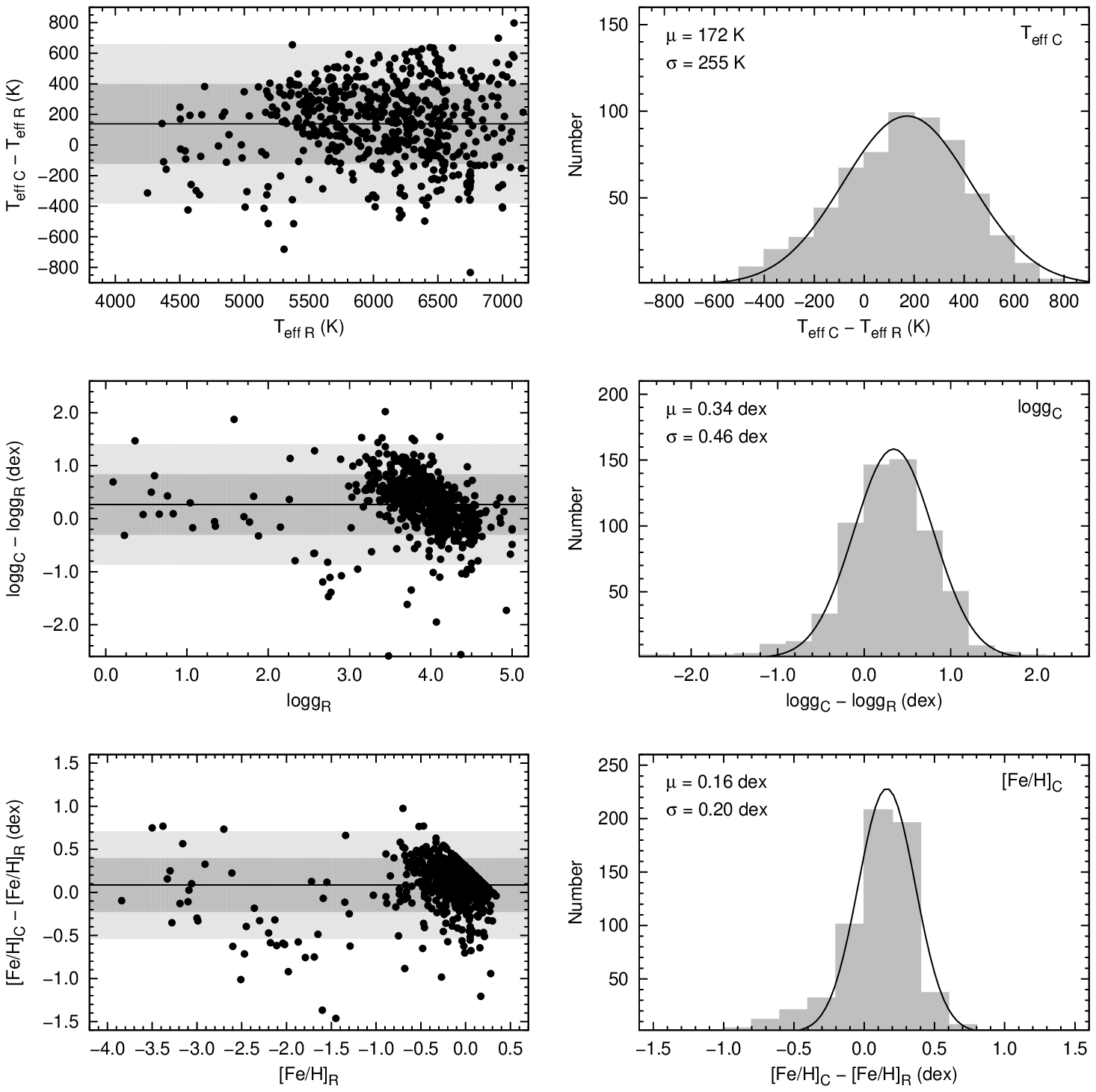}
\caption{Left panels: Differences between the (corrected) atmospheric parameters 
determined by the n-SSPP, Teff$_C$, logg$_C$, and [Fe/H]$_C$, and the
values from RAVE, Teff$_R$, logg$_R$, and [Fe/H]$_R$, reported in the
literature, as a function of the RAVE spectroscopic values. Filled
symbols refer to the program stars. The horizontal solid line is the
average of the residuals, while the darker and lighter shaded areas
represent the 1-$\sigma$ and 2-$\sigma$ regions, respectively. Right
panels: Histograms of the residuals between the corrected n-SSPP and
high-resolution parameters shown in the left panels. Each panel also
lists the average offset and scatter determined from a Gaussian fit.}
\label{atmMR}
\end{figure*}

Figure~\ref{atmMR} shows that the comparison of  Teff$_C$, logg$_C$, and
[Fe/H]$_C$ with the (non-suspect) RAVE determinations are significantly
worse for Teff$_C$, but commensurate with the comparisons to the
high-resolution results for logg$_C$ and [Fe/H]$_C$.  There are too few
stars in common between the stars with both RAVE parameter estimates and
high-resolution estimates to make meaningful comparisons.  

\citet{beers2014} also used literature values of [C/Fe], based on
high-resolution spectroscopic analyses, to derive corrections for the
n-SSPP estimates of [C/Fe], as:

\begin{equation}
{\rm [C/Fe]}_C    = {\rm [C/Fe]}_S - ( - 0.068 \cdot {\rm [C/Fe]}_S + 0.273)
\end{equation}

\noindent The corrected values are listed as [C/Fe]$_C$ in column (8) of
Table~\ref{tab3}. This table also lists, in column (9), the absolute
value of the carbon abundance, $A$(C) = log$\epsilon$\,(C)\footnote{$A$(C) 
is not measured directly, as it can be from high-resolution
spectroscopy, but rather, it is obtained from medium-resolution
determinations using $A$(C) = [C/Fe] + [Fe/H] + $A$(C) $_{\odot}$, where
we adopt the Solar abundance of carbon from \citealt{asplund2009},
$A$(C)$_{\odot}$ = 8.43.}. We assume, following Beers et al., that
external errors for [C/Fe]$_C$ are on the order of $\sim 0.20$ dex.

The parameter CEMP, shown in column (10) of Table~\ref{tab3}, indicates
whether the star is considered carbon enhanced: CEMP = ``C," satisfying
[C/Fe]$_C > +0.7$, CC $\ge 0.7$, and EQW $\ge 1.2$; of intermediate
carbon enrichment: CEMP=``I," satisfying $+0.5 < $ [C/Fe]$_C \le +0.7$, CC
$\ge 0.7$, and EQW $\ge 1.2$; or carbon normal: CEMP = ``N," satisfying
[C/Fe]$_C \le +0.5$, CC $\ge +0.7$, and EQW $\ge 1.2$. Stars with upper
limits on their carbon ratios are indicated with CEMP = ``U," (these
include stars with DETECT = ``U," CC $\ge +0.7$, and DETECT = ``D" but
CC $< +0.7$). Stars without carbon measurements are listed as CEMP
= ``X". There are 48 stars listed with CEMP = ``C," 29 with CEMP = ``I,"
1362 with CEMP = ``N," and 116 with CEMP = ``U".  


\begin{figure*}[!ht]
\epsscale{1.15}
\plotone{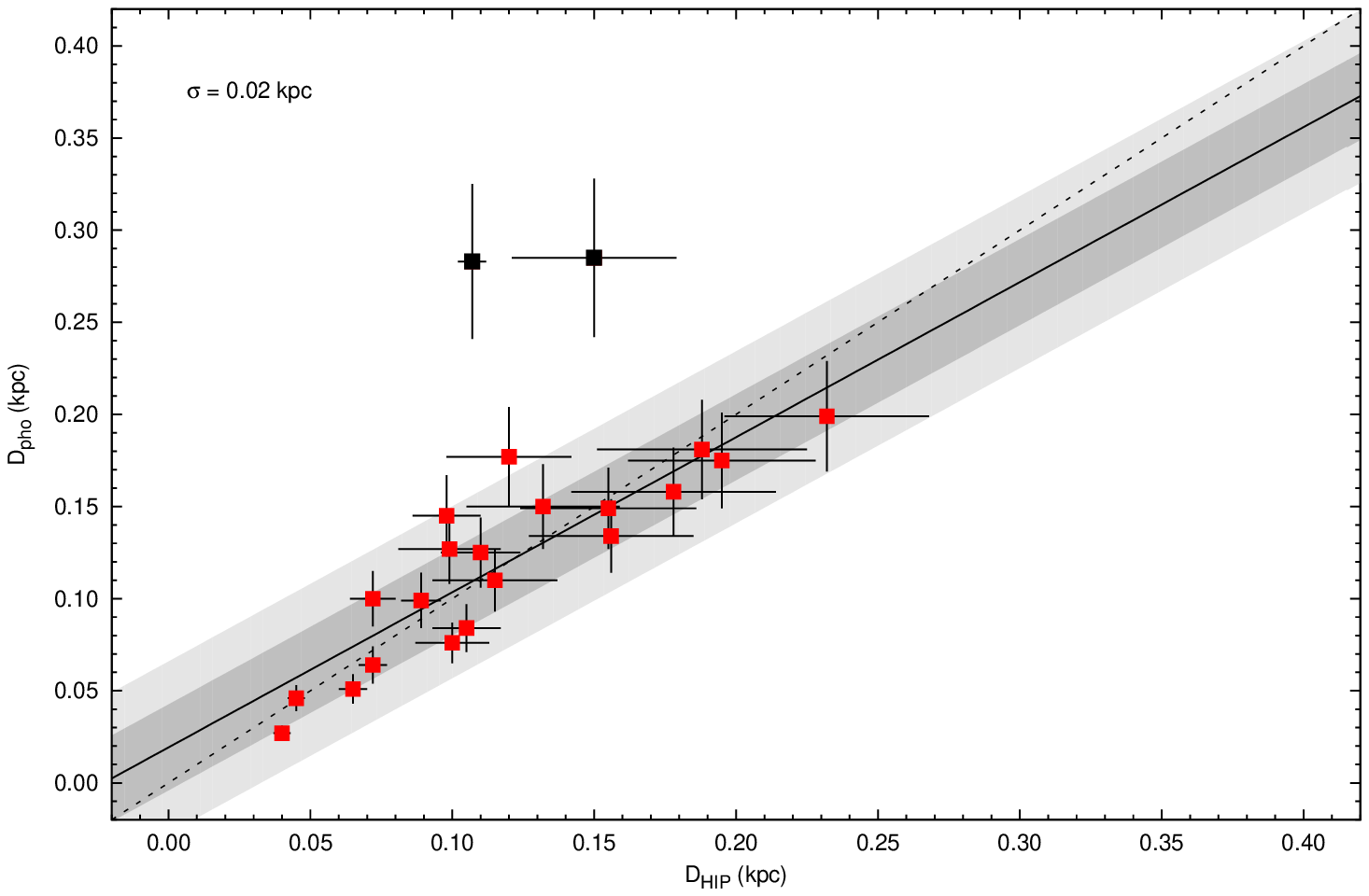}
\caption{Comparison of the photometrically estimated distances,
$D_{\rm pho}$, with the trigonometric distance estimates, $D_{\rm
 HIP}$, for stars with sufficiently accurate Hipparcos parallaxes
($\sigma_{\pi_{\rm HIP}}/\pi_{\rm HIP} \le 0.20$). The dashed line
is the one-to-one line, while the solid line is a robust regression
fit to the data. The darker and lighter shaded areas
represent the 1-$\sigma$ and 2-$\sigma$ regions about the linear
fit, respectively, based on a Gaussian fit to the residuals.
The most deviant stars include one giant and one dwarf.}
\label{dist}
\end{figure*}

\subsection {Distance Estimates and Proper Motions}

Distances to individual stars in our sample are estimated using the
$M_V$ vs. $(B-V)_0$ relationships described by \citet{beers2000}. These
relationships require that the likely evolutionary stage of a star be
given. Assignments to evolutionary stage, based on the derived
(corrected) stellar atmospheric parameters, are as follows: dwarf, D
(logg$_C \ge 4.0$), turnoff, TO (3.5 $\le $ logg$_C < 4.0$), subgiant or
giant, G (logg$_C < 3.5$). Note that refinements to this scheme,
designed to resolve the possible incorrect assignments of TO stars at
cooler temperatures, are adopted as described in \citet{beers2012}.
Following \citet{santucci2015}, stars with effective temperature Teff$_C
\ge 6000$~K and logg$_C \le 3.5$ are classified as field
horizontal-branch (FHB) stars.

Based on previous tests of this approach, we expect the distances
assigned as described above to be accurate to on the order of 15\%.
Fortunately, there are a small number of stars (24) in our sample with
reliable distance estimates available from Hipparcos parallax
measurements, listed in Table~\ref{tab4}, using the \citet{leeuwen2007}
reduction. Column (1) lists the star names, column (2) is the assigned
evolutionary type, column (3) is the Hipparcos parallax, $\pi_{\rm
HIP}$, column (4) is the error on parallax, $\sigma_{\pi_{\rm HIP}}$,
and column (5) is the ratio $\sigma_{\pi_{\rm HIP}/\pi_{\rm HIP}}$. In
order to be considered a reliable estimate of the parallax, this ratio
is required to be less than 0.20. The parallax distance estimate and
its error are listed as $D_{\rm HIP}$ and $\sigma_{\rm D_{\rm HIP}}$ in
columns (6) and (7), respectively. The estimated photometric distances,
$D_{\rm pho}$, and their errors, $\sigma_{D_{\rm pho}}$, are provided in
columns (8) and (9), respectively.

Figure~\ref{dist} presents a comparison of the distances calculated on the basis
of the photometric estimates and Hipparcos parallaxes. From inspection
of this figure, the great majority of the stars have commensurate
distance estimates; the one-sigma scatter of the residuals varies
between 10\% and 20\%, on the order of our adopted distance errors of
15\%. The most deviant stars include one giant and one dwarf.  

Table~\ref{tab5} lists star names in column (1), and the assigned type
classifications, photometric distance estimates, and their errors, in
columns (2)-(4), respectively. 

The great majority of our program stars (1732 stars) have reasonably
high-quality proper motions available from the UCAC4 catalog
\citep{zacharias2013}, the SPM4 catalog \citep{girard2011} (321 stars),
or the Hipparcos \citep{leeuwen2007} and Tycho~II catalogs
\citep{hog2000} (66 stars).  We have chosen to adopt, where possible, the UCAC4
proper motions, since they exist for almost all of our stars, and
generally have very small errors. The only exceptions are that, when
proper motions are available from the Hipparcos or Tycho II catalogs, we
adopt those. The final results are listed as $\mu_{\alpha}$ and
$\mu_{\delta}$, for the proper motions in the right ascension and
declination directions, respectively, in columns (5) and (6) of
Table~\ref{tab5}. Their associated errors are listed in columns (7) and
(8). The source of the adopted proper motion is listed in column (9).
Note that, for a small number of stars for which some ambiguity exists
as to which star of a listed pair is the one intended (generally those
with ``A,'' ``B,'' or ``F'' appended to their names), we did not adopt
any proper motions.

\begin{figure}[!ht]
\epsscale{1.15}
\plotone{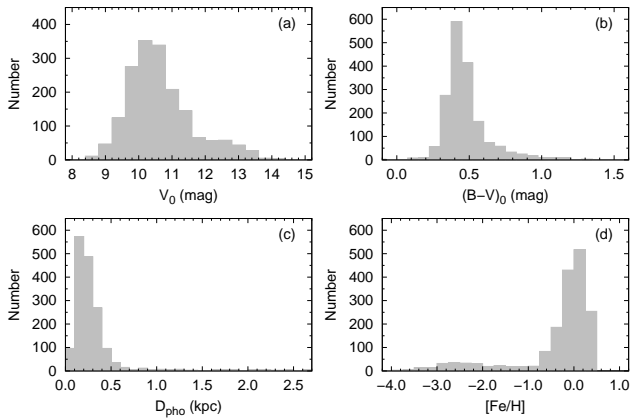}
\caption{Distributions of (a) absorption-corrected V$_0$ magnitudes, (b)
de-reddened $(B-V)_0$ colors, (c) photometric distance estimates, $D_{\rm pho}$,
and (d) metallicity estimates, [Fe/H], for our program stars.}
\label{sample}
\end{figure}

\section{A Kinematic Analysis of the Combined Frebel et al. (2006)
and Beers et al. (2014) Samples}\label{kinematics}

In this section we examine the kinematic properties of our program
stars, in combination with a similar local sample of stars originally
identified by \citet{bidelman1973} and discussed by \citet{beers2014}.
The stellar parameter estimates and derived kinematic quantities of this
latter sample were determined in an essentially identical manner as for
our program stars, and they supplement the numbers of stars with lower
metallicity for our subsequent analysis. The corrections to the
n-SSPP-derived atmospheric-parameter estimates and [C/Fe] for our
program stars are identical to those used by \citet{beers2014}. For
simplicity, we drop the ``C'' subscript on the corrected stellar
atmospheric parameters and the carbonicity estimates in the analysis
that follows, although it is understood that these are the quantities we
have adopted.

Figure~\ref{sample} shows the distribution of the absorption-corrected
$V_0$ magnitudes, de-reddened $(B-V)_0$ colors, distance estimates,
$D_{\rm pho}$, and estimates of metallicities, [Fe/H], for our program
stars from Paper~I. As is immediately clear from inspection of this
figure, this is a very local sample of stars, with $\sim 90$\% of the
stars located within 0.5 kpc of the Sun. Although the majority of the
sample stars have metallicities close to Solar, some 20\% (351 stars) of
the stars with available metallicity estimates have [Fe/H] $\le -0.5 $,
14\% (248 stars) have [Fe/H] $\le -1.0$, 12\% (213 stars) have [Fe/H]
$\le -1.5$, and 10\% (171 stars) have [Fe/H] $\le -2.0$. Figure 6 of
\citet{beers2014} shows similar information for that sample.  As can be
appreciated from inspection of that figure, these stars include a
larger fraction of giants, which explore slightly farther from the Sun,
up to 2 kpc (although $\sim$90\% are within 1 kpc of the Sun). Unlike
the Paper~I stars, almost half of the supplemental sample (145 stars) have
[Fe/H] $\le -1.0$; there are also 36 stars with [Fe/H] $\le -2.0$),
which makes them useful for our exploration of the metal-poor
populations of the Galaxy.

\subsection {Determination of $U$,$V$,$W$ Velocity Components and Orbital Eccentricities
for the Frebel et al. (2006) Sample}

The derivation of space motions and orbital parameters of our program
stars from Paper~I follows the procedures described by
\citet{carollo2010}, which for convenience are summarized below. Similar
procedures were employed by \citet{beers2014} for the supplemental
stars; results are listed in Table 5 of that paper.

Corrections for the Solar motion with respect to the Local Standard
of Rest (LSR) are applied during the course of the calculation of the
full space motions; here we adopt the values $(U,V,W)$=(9, 12, 7) \kms\
\citep{mihalas1981}. We follow the convention that $U$ is
positive in the direction away from the Galactic center, $V$ is positive
in the direction of Galactic rotation, and $W$ is positive toward the
north Galactic pole. It is also convenient to obtain the rotational
component of a star's motion about the Galactic center in a cylindrical
frame, denoted as \vphi, and calculated assuming that the LSR is on a
circular orbit with a value of 220 \kms\ \citep{kerr1986}. Our assumed
values of the Solar radius (\rsun\ $= 8.5$ kpc) and the circular
velocity of the LSR are both consistent with two recent independent
determinations of these quantities by \citet{ghez2008} and
\citet{koposov2009}. \citet{bovy2012} obtained an estimate of the Milky
Way's circular velocity at the position of the Sun of V$_c$(\rsun) =
$218 \pm 6$ \kms, based on an analysis of high-resolution spectroscopic
determinations from APOGEE, which is also consistent with our adopted
value.

The orbital parameters of the stars, including the perigalactic distance
(the closest approach of an orbit to the Galactic center), \rperi, the
apogalactic distance (the farthest extent of an orbit from the Galactic
center), \rapo, of each stellar orbit, and the orbital eccentricity,
$e$, defined as $e$ = (\rapo~$-$~\rperi)/(\rapo~$+$~\rperi), as well as
\zmax\ (the maximum distance that a stellar orbit achieves above or
below the Galactic plane), are derived by adopting an analytic
St\"ackel-type gravitational potential \citep[which consists of a
flattened, oblate disk, and a nearly spherical massive dark-matter halo;
see the description given by][ Appendix A]{chiba2000}, and integrating
their orbital paths based on the starting point obtained from the
observations.  

Table~\ref{tab6} provides a summary of the above calculations. Column
(1) provides the star names. Columns (2) and (3) list the positions of
the stars in the meridional ($R,Z$)-plane. The derived $U$,$V$,$W$ velocity
components are provided in columns (4)-(6); their associated errors are
listed in columns (7)-(9). Column (10) lists the velocity projected onto
the Galactic plane ($V_R$, positive in the direction away from the
Galactic center), while column (11) lists the derived rotation velocity,
\vphi. The derived \rperi\ and \rapo\ are given in columns (12) and
(13), respectively. Columns (14) and (15) list the derived \zmax\ and
orbital eccentricity, $e$, respectively.  The INOUT parameter listed in 
column (16) is set to 1 if the star is considered in our kinematic
analysis, and set to 0 if not.

\begin{figure}[!ht]
\epsscale{1.15}
\plotone{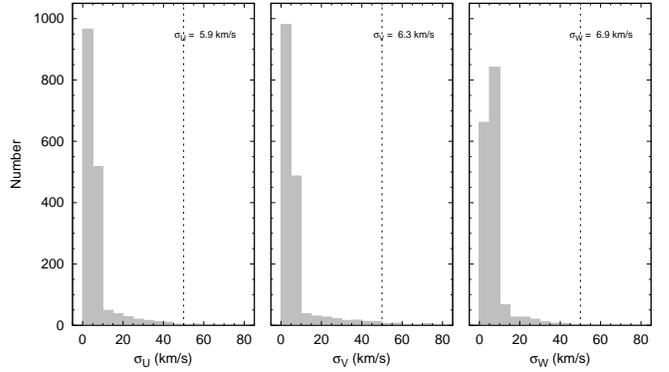}
\caption{Errors in the estimation of the local velocity components of
the space motions for the Paper~I stars.
The vertical dashed lines at 50 \kms\ indicate the maximum
individual errors allowed for a given star to be included in the
subsequent kinematic analysis.  The legends provide the mean errors for
the accepted stars.}
\label{sigma}
\end{figure}

Errors on our derived estimates of the individual components of the
space motions take into account an estimated 15\% error in the
photometric distances, as well as the individual errors in the proper
motions (average errors on our adopted proper motions is 1.3
mas~yr$^{-1}$ in each of the RA and DEC component directions), and in the
adopted radial velocities (2 \kms\ for the high-resolution
determinations, 5 \kms\ for the moderate-resolution determinations, and
10 \kms\ for the medium-resolution determinations). Figure~\ref{sigma}
shows the distributions of these errors. After removing the 145 
stars that are missing one or more of the input quantities used for the
determination of their space motions, or with individual estimated
errors in any one of the three components of space motion larger than 50
\kms, the average errors for our program sample are $\sigma (U, V, W)$ =
(5.9, 6.3, 6.9) \kms. These are slightly lower errors than were achieved
for the supplemental stars from \citet{beers2014} (after removing stars
having errors in $U$, $V$, or $W$ greater the 50 \kms), who reported
$\sigma (U, V, W)$ = (7.9, 9.1, 6.5) \kms, presumably due to the
inclusion of more distant stars with less certain distances and proper
motions.

For the remaining analysis, we combine our program stars from Paper~I
with the supplemental sample based on the \citet{beers2014} analysis of
the `weak metal' stars from \citet{bidelman1973}. For the purpose of the
kinematic analysis, both samples have had stars with errors in any of
the $U,V,W$ velocity components in excess of 50 \kms\ removed from
consideration.

\subsection{Distributions of $U,V,W$, and \zmax\ vs. [Fe/H]}

\begin{figure}[!ht]
\epsscale{1.15}
\plotone{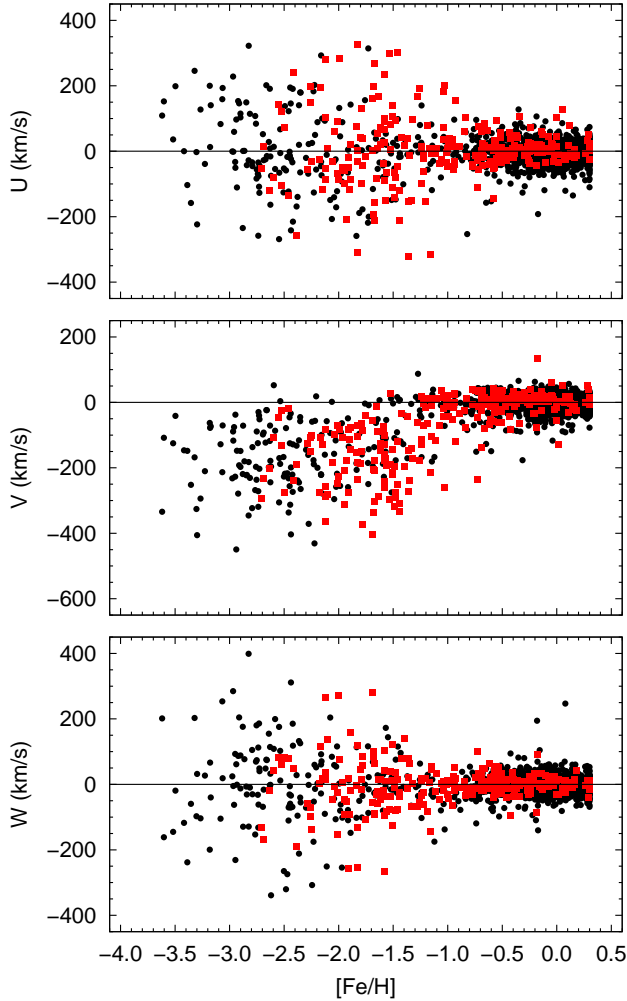}
\caption{Local velocity components for the combined sample of Paper~I stars
(shown as black dots) and the supplemental stars from \citet{beers2014}
(shows as red squares) with available $UVW$ estimates, as a function of
metallicity, [Fe/H]. Note the existence of stars with low velocity
dispersions in their estimated components down to at least [Fe/H] =
$-1.3$, and possibly a little lower. Stars with errors in any of the
individual derived components of motion exceeding 50 \kms\ are excluded.
}
\label{uvw}
\end{figure}

Figure~\ref{uvw} presents the individual components of the space
motions, as a function of [Fe/H], for our combined sample with accepted
kinematic estimates; the program stars from Paper~I are shown as black
dots, while the supplemental-sample stars are indicated as red squares.
From inspection of this figure, the two samples cover
similar ranges of [Fe/H], although in different proportion -- the
Paper~I sample dominates above [Fe/H] = $-1.0$, the supplemental-sample
stars exceed the Paper~I stars in the metallicity interval $-2.0 <$
[Fe/H] $< -1.0$ by about a factor of two, and the Paper~I stars dominate
the combined sample of stars with [Fe/H] $< -2.0$, in particular for
[Fe/H] $< -3.0$. The combined sample is heavily populated by stars in
the thin-disk and thick-disk stellar populations. Some low-metallicity
stars with $V$ velocities in the interval $-$40 to $-$80 km s$^{-1}$ are
also present, and are likely associated with the MWTD.

\begin{figure}[!ht]
\epsscale{1.20}
\plotone{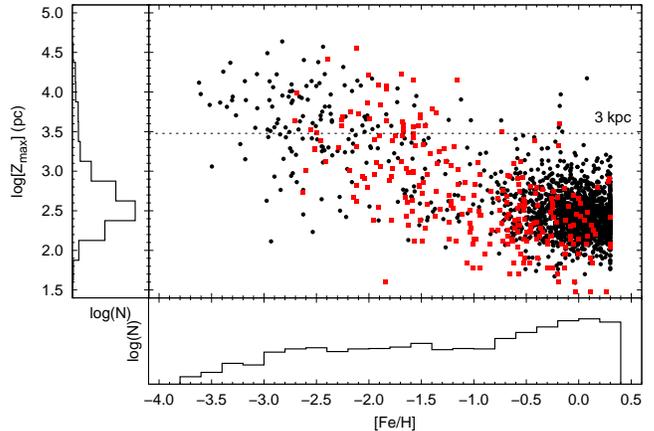}
\caption{Distribution of $Z_{\rm max}$, the largest distance above or below
the Galactic plane achieved by a star during the course of its orbit, as
a function of metallicity, [Fe/H], for the combined sample of Paper~I stars
(shown as black dots) and the supplemental stars from \citet{beers2014}
(shows as red squares). The marginal distributions of each variable are
shown as histograms. The horizontal dashed line provides a reference at
3 kpc. Very few stars with metallicity [Fe/H] $ > -1.5$ achieve orbits
that reach higher than this location. Note the logarithmic scale for
$Z_{\rm max}$. Stars with errors in any of the individual derived
components of motion exceeding 50 \kms\ are excluded. }
\label{zmax}
\end{figure}

Figure~\ref{zmax} is a plot of \zmax, as a function of [Fe/H], for the
combined sample of stars. From inspection of this figure, it is clear
that both the Paper~I and supplemental samples explore similar regions
of this space, which further justifies carrying out a joint kinematic
analysis. For the remainder of our analysis, we thus choose to suppress
identification of the individual samples. 

As seen in Figure~\ref{zmax}, only a handful of stars with metallicities
above [Fe/H] = $-1.5$ are found with \zmax\ $> 3 $ kpc. Following
previous results from, e.g., \citet{carollo2010}, stars with \zmax\ $\le
3 $ kpc and $-1.8 \le $ [Fe/H] $\le -0.8$ are likely to be associated
with the MWTD, although some overlap with the inner-halo population is
not precluded, especially at the low end of this metallicity range.
Further interpretation of the nature of the MWTD as an individual
component is limited by the small numbers of stars, even in
the combined sample, that are available in the pertinent metallicity
interval.

\subsection{The [Fe/H] vs. Eccentricity Diagram}

\begin{figure}[!ht]
\epsscale{1.20}
\plotone{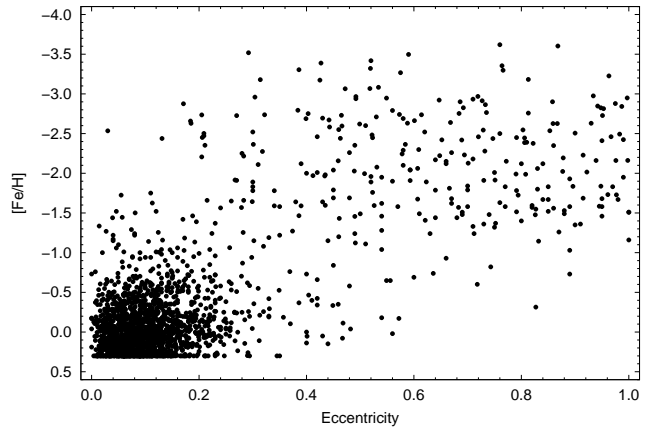}
\caption{Distribution of metallicity, [Fe/H], for the combined sample of stars, as a
function of derived orbital eccentricity. Stars with errors in any of
the individual derived components of motion exceeding 50 \kms\ are
excluded. }
\label{efeh}
\end{figure}

Figure~\ref{efeh} shows a plot of [Fe/H], as a function of orbital
eccentricity, for the combined sample of stars. As has been seen
previously (e.g., \citealt{norris1985}; \citealt{chiba2000}; 
\citealt{carollo2007}; \citealt{carollo2010}; \citealt{beers2014}), 
the distribution of orbital eccentricity for these {\it
non-kinematically-selected} stars exhibits a very broad metallicity
distribution, outside of the region of the metal-richest stars with $e
\le 0.2-0.3$, as expected from the currently favored hierarchical
assembly model for the formation of the Milky Way. 

\vspace{1cm}

\subsection{The Toomre Diagram, Distribution of \vphi, Integrals of Motion, and the Lindblad
Diagram}

The so-called Toomre diagram (a plot of ($U^2 + W^2)^{1/2}$, the
quadratic addition of the $U$ and $W$ velocity components, as a function
of the rotational component, $V$), the distribution of orbital rotation
velocity, \vphi, for cuts in orbital eccentricity and [Fe/H], plots of
the perpendicular angular momentum component, L$_{\bot}$, as a function
of the vertical angular momentum component, L$_Z$, and the Lindblad
diagram (a plot of the integral of motion representing the total energy,
E, as a function of L$_Z$) are commonly used to investigate the nature
of the kinematics of stellar populations in the Galaxy. Given the high
quality of the estimated kinematics for our combined sample of stars, it
is worthwhile to investigate what can be learned from inspection of
these diagrams, as discussed individually below.

\subsubsection{The Toomre Diagram}

\begin{figure}[!ht]
\epsscale{1.2}
\plotone{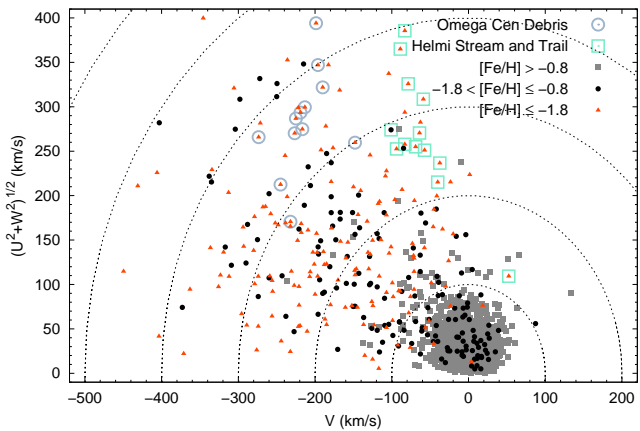}
\caption{Toomre diagram of ($U^2 + W^2)^{1/2}$ vs. $V$ for
stars in the combined sample with available $UVW$ velocity components,
in three regimes of metallicity as indicated in the legend. The legend
also indicates the color/symbol coding used to indicate likely members
of stars in the debris stream associated with the globular cluster
$\omega$ Cen (light-blue circles) and the Helmi et al. stream/trail
(light-green squares). See text for more details. Note the presence of
the intermediate-metallicity ($-1.8 < $ [Fe/H] $\le -0.8$) stars both
inside and outside the region with low ($U^2 + W^2)^{1/2}$ and high
orbital rotation velocities (($U^2 + W^2)^{1/2} \lesssim 100$ \kms,
$-100 < V < 100$ \kms). Stars with errors in any of the individual derived components
of motion exceeding 50 \kms\ are excluded. }
\label{toomre}
\end{figure}

Figure~\ref{toomre} shows the Toomre diagram for the combined sample of
stars; the legend indicates the metallicity intervals chosen to roughly
separate stars expected to belong to the thick (or thin) disk ([Fe/H] $>
-0.8$), the MWTD ($-1.8 < $ [Fe/H] $\le -0.8$), and the halo system
([Fe/H] $\le -1.8$), taking our guidance in selecting these intervals
from \citet{carollo2010}. As expected, the more metal-rich stars in both
samples are primarily found in the region with low ($U^2 + W^2)^{1/2}$
and high orbital-rotation velocities, $(U^2 + W^2)^{1/2} \lesssim 100$
\kms, $-100 < V < 50$ \kms, while stars with intermediate metallicities are 
divided between those inside and outside this region. We expect that
many of the intermediate-metallicity stars inside this region are
associated with the MWTD component. It is also clear from inspection of
this figure that the lowest metallicity stars, with [Fe/H] $\le -1.8$,
are the dominant contributors to the distribution of stars in the
higher-energy regions (those beyond the circle that intersects $V =
-300$ \kms), as might be expected if they primarily comprise members of
the outer-halo population, with some overlap from members of the
inner-halo population. The stars with energies that place them between
the $V = -300$~\kms\ and $V = -200$~\kms\ surfaces exhibit a broader
range of metallicity, as expected from overlapping inner- and outer-halo
populations.

\begin{figure*}[!ht]
\epsscale{1.05}
\plotone{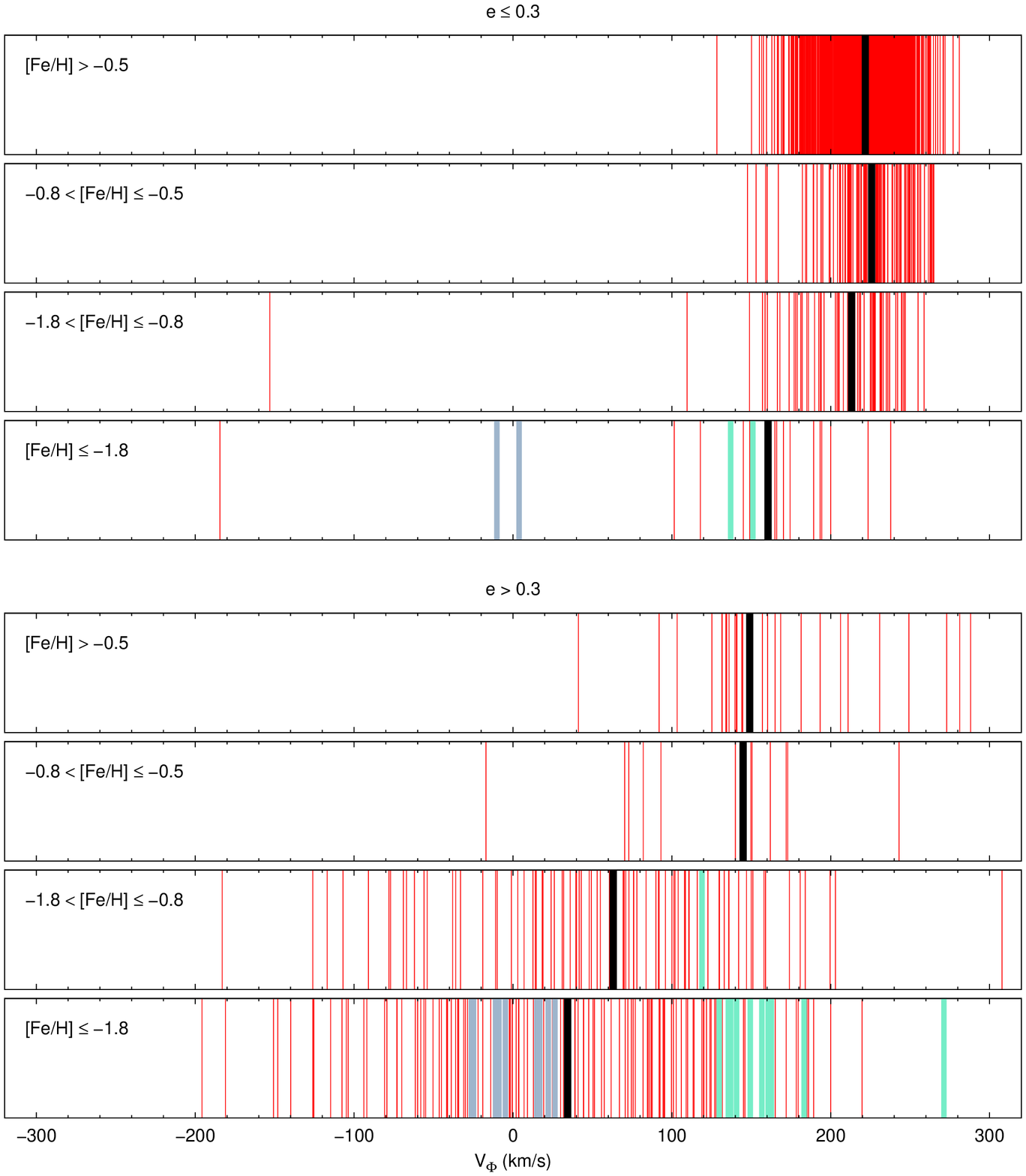}
\caption{Stripe-density diagrams of the rotational velocity, \vphi, 
for stars in the combined sample. The plots are split into
low-eccentricity ($e \le 0.3$; upper panels) and high-eccentricity ($e >
0.3$; lower panels) sub-samples. Each sub-sample is further divided into
metallicity intervals chosen to separate regions dominated by individual
components of the disk and halo systems. See text for more details. The
black stripes indicate the mean \vphi\ for stars in each subset. The
light-blue and light-green stripes indicate stars identified as likely
members of stars in the debris stream associated with the globular
cluster $\omega$ Cen and the Helmi et al. stream/trail, respectively. Stars
with errors in any of the individual derived components of motion
exceeding 50 \kms\ are excluded.}
\label{vphi_stripe}
\end{figure*}

Figure~\ref{toomre} also indicates two subsets of (newly-identified)
stars in the combined sample that may belong to previously identified 
structures in phase-space: (1) Likely members of the stream/trail of
stars first identified by \citet{helmi1999} and further populated by stars in the sample
considered by \citet{chiba2000}, indicated by light-green squares, and (2)
Possible members of the debris stream associated with the globular
cluster $\omega$ Cen, following the work of \citet{dinescu2002},
\citet{klement2009}, and \citet{majewski2012}, indicated by light-blue circles. 
Justification for the selection of these stars is provided below.

\subsubsection{Distribution of \vphi}

Figure~\ref{vphi_stripe} is a stripe-density diagram of the distribution of
\vphi\ for our combined sample of stars, for metallicity intervals chosen to
emphasize the various kinematic components of the Milky Way, split into
two regions of orbital eccentricity, $e \le 0.3$ (upper panels; expected
to be dominated by members of disk system) and $e > 0.3$ (lower panels;
expected to be dominated by members of the halo system). For each
interval in metallicity, the black stripes indicate the mean \vphi\ for
that sub-sample of stars. The light-green and light-blue stripes
indicate stars that we argue below are candidate members of the Helmi
et al. stream/trail and the $\omega$ Cen debris streams, respectively.

Inspection of Figure~\ref{vphi_stripe} generally meets with expectation, 
based on previous work. The low-eccentricity stars for all three
sub-panels with [Fe/H] $> -1.8$ exhibit rotational properties consistent
with the disk system of the Milky Way (thin disk, thick disk, and MWTD)
while those with [Fe/H] $\le -1.8$ appear to be primarily members of the
inner- and outer-halo populations. The high-eccentricity stars
preferentially populate the sub-panels with $-1.8 < $[Fe/H] $\le -0.8$
and [Fe/H] $\le -1.8$, consistent with membership in the inner- and
outer-halo populations, with overlapping contributions from each.     

It is worth noting that the presence of the putative members of the 
two debris streams has a potentially large impact on interpretation of
the distribution of \vphi\ among the high-eccentricity stars with [Fe/H]
$< -1.8$, populating both the central region of the stripe plot (Helmi et al.
stream/trail) as well as the high-velocity tail ($\omega$ Cen debris
stream).
  
\begin{figure*}[!ht]
\includegraphics[width=08.8cm]{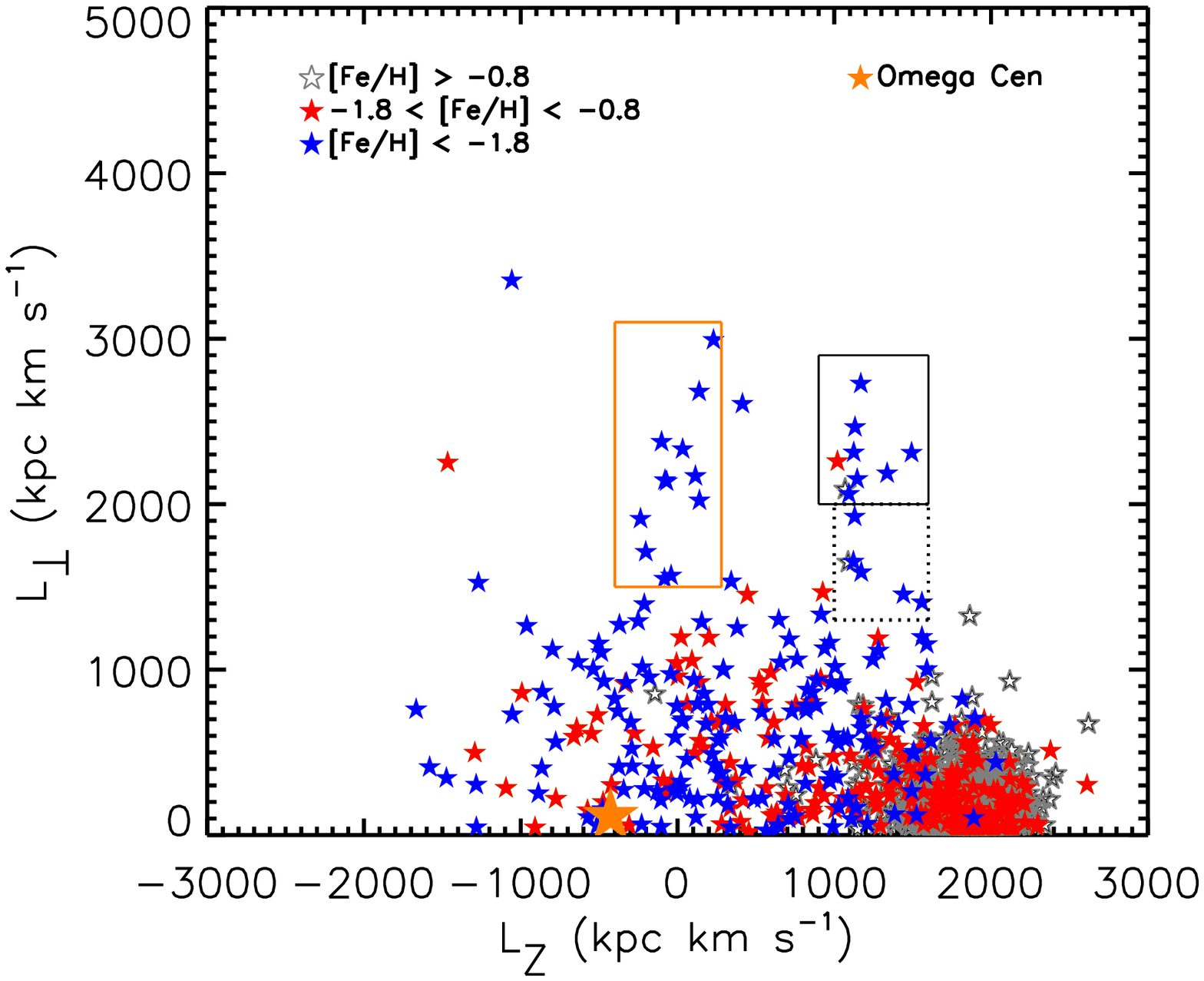}
\includegraphics[width=10.0cm]{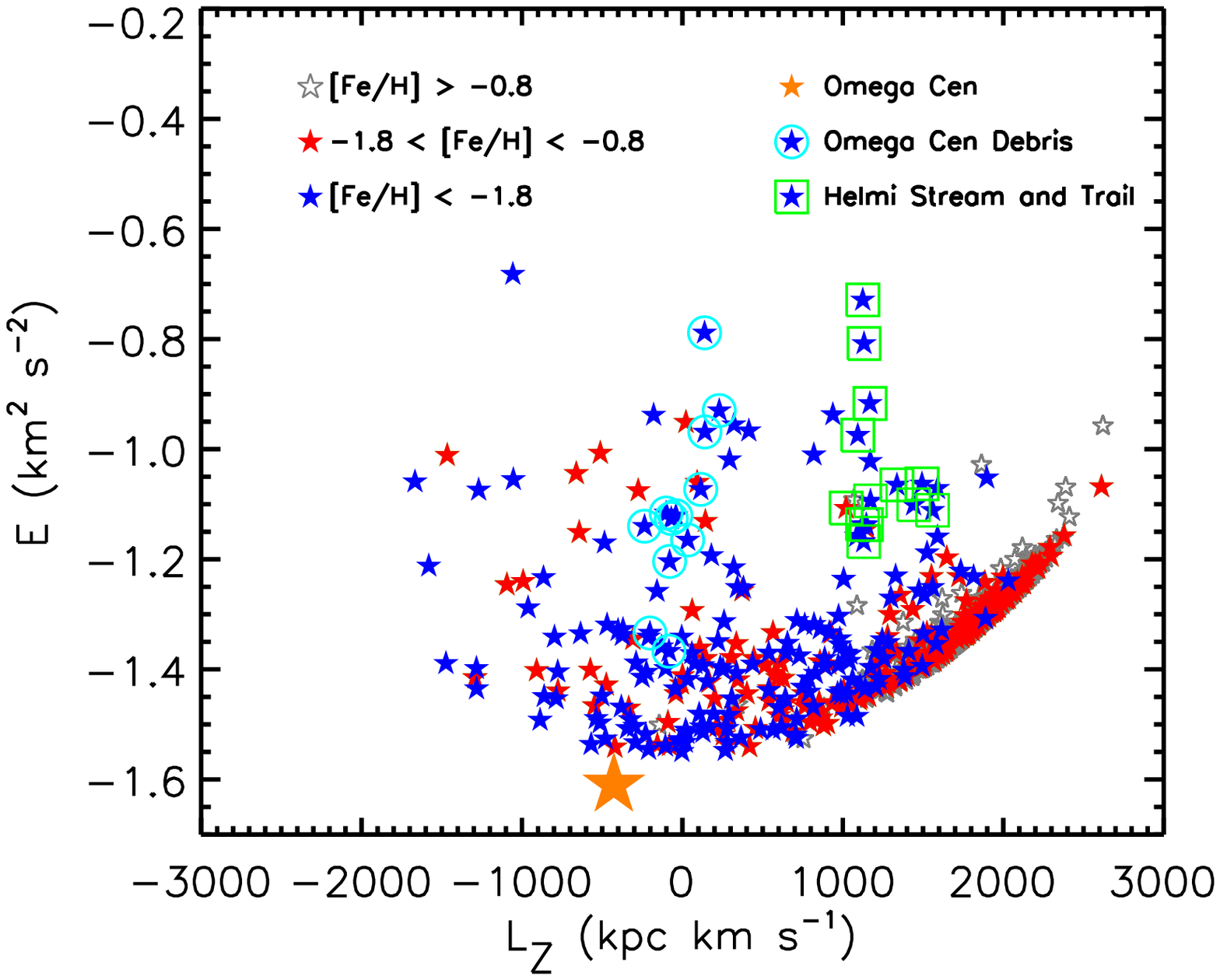}
\caption{Left panel: Distribution of the angular momentum components L$_{\bot}$, L$_{Z}$, for the
combined sample of stars, over three ranges of metallicity as shown in
the legend. The solid and dotted black boxes denote the region of the
clumps that are likely associated with the Helmi et al. stream and
trail, respectively. The orange box represents the region of the
putative debris stream associated with the $\omega$ Cen globular
cluster. The position of this cluster in this diagram is indicated with
the large orange star. Stars with errors in any of the individual
derived components of motion exceeding 50 \kms\ are excluded. 
Right panel: Lindblad diagram of the distribution of the total energy, E (in units of 10$^{5}$), as a function
of the vertical angular momentum, L$_{Z}$, over three ranges of
metallicity as shown in the legend. Likely members of the Helmi et al.
stream and its trail are highlighted with light-blue circles; stars that
are likely members of the putative $\omega$ Cen debris stream are
indicated by light-green squares. The position of this cluster in this
diagram is indicated with the large orange star. Stars with errors in
any of the individual derived components of motion exceeding 50 \kms\
are excluded.}
\label{amomenta}
\end{figure*}


\subsubsection{L$_{\bot}$ vs. L$_Z$}

The left-hand panel of Figure~\ref{amomenta} shows the distribution of stars in angular
momentum space (L$_{\bot}$, L$_{Z}$), where L$_{\bot}$ = (L$_{X}$$^2 +
$L$_{Y}$$^2$), and L$_{Z}$ is the vertical angular momentum. The three
different ranges of metallicity are identified with different colors,
shown in the figure legend.    

Several interesting features are seen in this diagram: (1) A clump of
stars with [Fe/H] $< -1.8$ (with the exception of two stars with higher
metallicity) located at L$_{\bot}$ $\sim$ 2000-2900 kpc km s$^{-1}$ and
L$_{Z}$ $\sim$ 800-1600 kpc km~s$^{-1}$ (identified by the solid black
box in the figure), and (2) The elongated distribution of stars with [Fe/H]
$\le -1.8$ located at L$_{\bot}$ $>$ 1500 km s$^{-1}$ and $-$400
$\lesssim$ L$_{Z}$ $\lesssim$ 300 kpc km s$^{-1}$ (indicated by the orange
box). 

The first feature was identified by \citet{helmi1999}, comprising 7
stars with [Fe/H] $\le -1.6$ and 12 stars with [Fe/H] $\le -1.0$.
\citet{chiba2000} detected the same stream among their sample of 1203
stars over similar ranges in metallicity.
They also identified a possible trail in angular momentum space located
at 1250 kpc km s$^{-1}$ $<$ L$_{\bot}$ $<$ 2000 kpc km s$^{-1}$ and 1200
kpc km s$^{-1}$ $<$ L$_{Z}$ $<$ 2000 kpc km s$^{-1}$, covering similar
metallicity ranges (their Figure 15). This is similar to the trail
identified in Figure~\ref{amomenta}, occupying the region defined by the
dotted black box, covering angular momentum ranges
L$_{\bot}$: [1300, 2000] kpc km s$^{-1}$, L$_{Z}$: [1000, 1600] kpc km
s$^{-1}$, but at lower metallicities, [Fe/H] $\le -1.8$. Note that a few
stars with metallicities above [Fe/H] = $-$1.8 are also within the areas
delimited by the two boxes associated with the Helmi et al.
stream/trail.

The second feature (orange box) is similar to the excess of stars
located in the phase-space noted by \citet{dinescu2002} within the
\citet{chiba2000} dataset, who argued that these stars may be part of a
debris stream associated with the globular cluster $\omega$
Cen. \citet{dinescu2002} found that most of the stars in this region
possessed slightly retrograde orbits, as is also the case for $\omega$
Cen, and another two clusters (NGC~362 and NGC~6779) that present similar
retrograde orbits. These authors also suggested that the cluster
$\omega$ Cen (shown as a large orange star in the figure), as well as
the two other globular clusters, may have been stripped, along with
numerous other stars, from a proposed parent dwarf galaxy, now 
dissolved into the halo-system population. 

\subsubsection{The Lindblad Diagram, E vs. L$_Z$}
 
The right-hand panel of Figure~\ref{amomenta} is the so-called Lindblad diagram for the combined
sample, split into the same metallicity ranges as in the left-hand panel.
Stars associated with the \citet{helmi1999}
stream and its trail are indicated with light-green boxes around them, while
those identified as possible members of the $\omega$ Cen debris stream
are indicated with light-blue circles around them. The Helmi et al. stream and
its trail occupy a range of orbital energy E: [$-$1.2, $-$0.7] km$^{2}$
s$^{2}$ (in units of 10$^{5}$), while the putative $\omega$ Cen stellar
debris stream stars have orbital energies spanning E: [$-$1.35, $-$0.8]
km$^{2}$ s$^{2}$. 

The stars we identify as members of these structures are listed in
column (1) of Table~\ref{tab7}, along with their coordinates (column 2),
photometry (columns 3 and 4), derived metallicity,
[Fe/H] (column (5), carbonicity, [C/Fe] (column 6), and absolute carbon
abundance, A$(C)$ (column (7), as well as their integrals of motion
(columns 9-11). We have verified that these stars are not among those
previously identified by \citet{chiba2000}. There are five CEMP stars
among the proposed $\omega$ Cen debris stream listed in this table, with
carbonicities in the range [C/Fe]: [$+$0.73,+1.47]. The listed absolute
carbon abundances for four of these stars, $A$(C), are all below 7.1;
according to the Yoon-Beers diagram of $A$(C) vs. [Fe/H] (\citealt{yoon2016}; Figure 1), they
would be classified as CEMP-no stars. There is one star in the proposed
$\omega$ Cen debris stream with $A$(C) $>$ 7.1, which would suggest its
identification as a CEMP-$s$ star. The CEMP sub-classifications are
shown in column (8) of Table~\ref{tab7}. 

In a previous study, \citet{majewski2012} identified a number of
carbon-enhanced stars from the Grid Giant Stream Survey sample that may 
be associated with the purported $\omega$ Cen debris stream.
Many of these stars exhibit enhanced [Ba/Fe] ratios, similar
to examples of the CEMP-$s$ stars previously identified in the cluster.
Given the relative rarity of CEMP stars found in most globulars, they
considered this compelling evidence that the field stars they identified
were indeed once associated with $\omega$ Cen. The one CEMP$-s$ and
four CEMP-no stream stars in our sample all have [Fe/H] $< -$2, falling
below the lower metallicity range associated with $\omega$ Cen
\citep{frinchaboy2002}. As indicated in the table, 14 of the listed
stars have existing high-resolution spectroscopy (most unpublished,
from our group).  We are in the process of obtaining high-resolution
spectroscopy for the stars in this table that presently lack this
information; the full sample will be described in due course.

\section{Carbon-Enhanced Metal-Poor Stars in the Combined Sample}
\label{cempsec}

\begin{figure}[!ht]
\epsscale{1.20}
\plotone{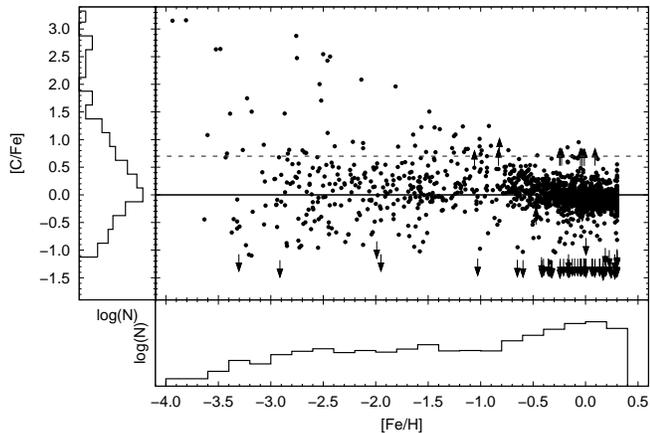}
\caption{Carbonicity, [C/Fe], as a function of the metallicity, [Fe/H], for 
the combined sample of stars with available measurements. Downward
arrows indicate derived upper limits for [C/Fe], and upward arrows
indicate lower limits. The marginal distributions of each variable are
shown as histograms. The horizontal dashed line marks the level of
carbon enhancement used in this paper to indicate CEMP stars, [C/Fe] $>
+0.7$. }
\label{cfe}
\end{figure}

Figure~\ref{cfe} shows the distribution of carbonicity, [C/Fe], as a
function of [Fe/H], for the stars in the combined sample. The general
increase in the level of [C/Fe] with decreasing [Fe/H] is clearly
evident, as is the increase in the frequency of CEMP stars with
decreasing metallicity below [Fe/H] $= -1.0$, as has been seen in
numerous previous studies. Note that our present study, following most
recent work, employs the criterion [C/Fe] $> +0.7$ (rather than [C/Fe]
$> +1.0$ as commonly used previously) to identify CEMP stars. It is
interesting to note the similarity of this figure to that reported by
\citet{rossi1999} (their Figure 2), which made these same points (as did
\citealt{norris1997}) almost twenty years ago. 

One early claim on the existence of an increased fraction of CEMP stars
with [Fe/H] $\le -2.0$ was made by Paper~I (9\%), somewhat lower than the
fractions reported by \citet{beers1992} ($\sim$ 14\%) and by authors of
other contemporaneous studies \citep[e.g.,][]{beers2005,cohen2005,
marstellar2005,lucatello2006}, on the order of $\sim$ 15 to 20\%. All
such estimates were, however, based on relatively small samples, and did
not account for the depletion of carbon for stars in advanced
evolutionary stages. 

\begin{figure}[!ht]
\epsscale{1.20}
\plotone{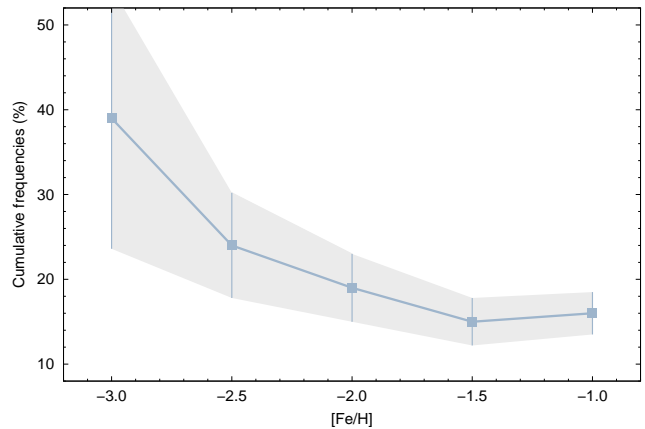}
\caption{Cumulative frequencies of CEMP stars, as function of
metallicity, [Fe/H], for stars in the combined sample with available
measurements.  A total of 328 stars with [Fe/H] $< -1.0$ are included in this diagram, 52 of
which are considered CEMP stars, with [C/Fe] $> +0.7$.  The error bars
shown are based on Poisson statistics.}
\label{frequencies}
\end{figure}

Figure~\ref{frequencies} shows the distribution of cumulative
frequencies for CEMP stars in our combined sample as a function of [Fe/H]. Although
the total numbers of CEMP stars in our sample is still small (N = 52), compared to
more recent work \citep[e.g.,][]{lee2013,placco2014}, the
behavior is similar. For completeness, we note that we obtain 
cumulative frequencies of CEMP stars of 19 $\pm$ 4\% for stars with
[Fe/H] $\le -2.0$, 24 $\pm$ 6\% for stars with [Fe/H] $\le -2.5$, and 39
$\pm$ 15\% for stars with [Fe/H] $\le -3.0$. These numbers compare well
with the cumulative frequencies of CEMP stars as a function of
decreasing metallicity from \citet{lee2013}, but are somewhat lower
than the frequencies reported by \citet{placco2014}, an analysis that
was based exclusively on stars with results from high-resolution
spectroscopic analyses, in particular if one considers their results
after corrections for the depletion of carbon in more-evolved stars.

\begin{figure*}[!ht]
\epsscale{1.025}
\plotone{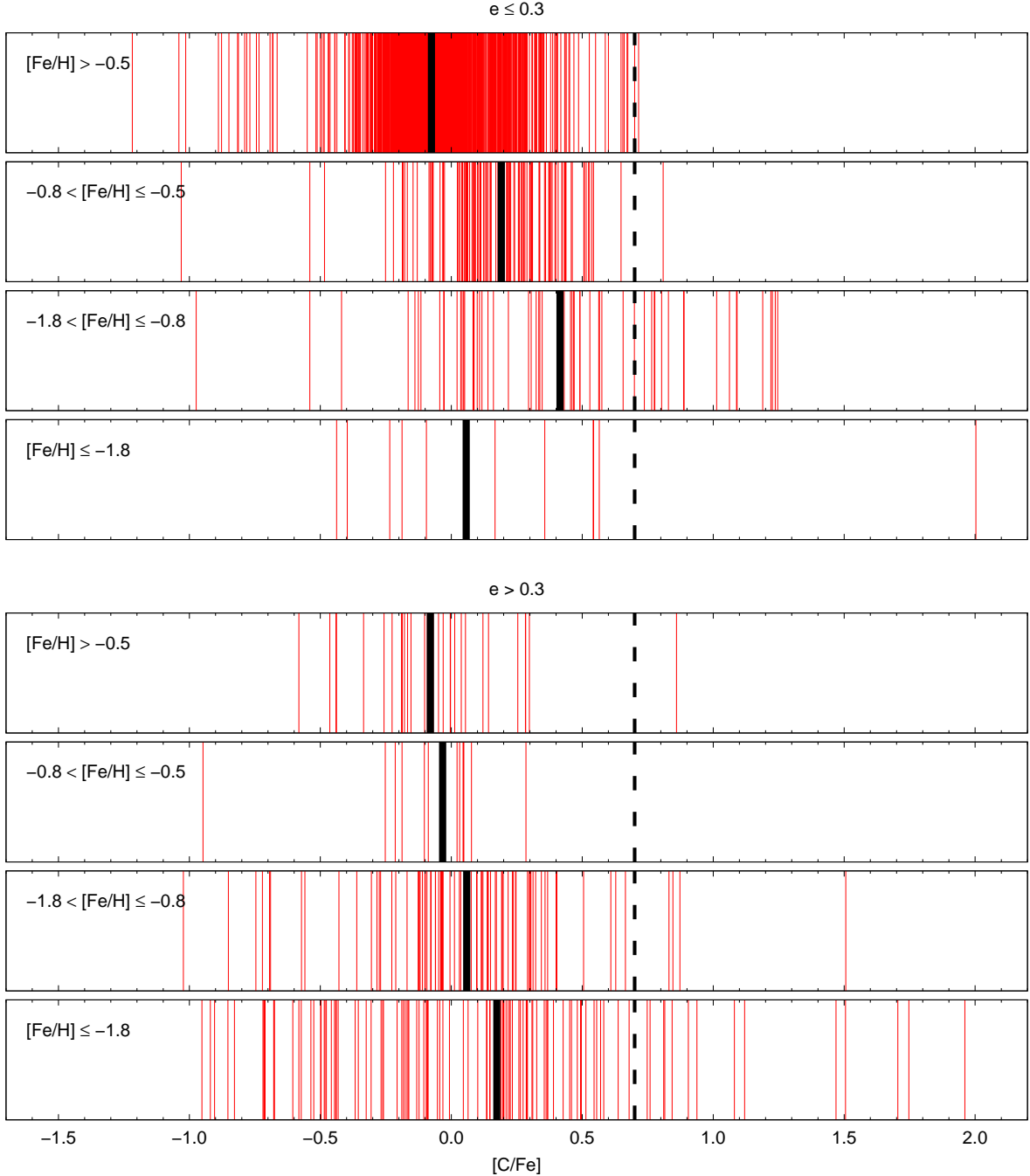}
\caption{Stripe-density diagrams of the carbonicity, [C/Fe], 
for stars in the combined sample. The plots are split into
low-eccentricity ($e \le 0.3$; upper panels) and high-eccentricity ($e >
0.3$; lower panels) sub-samples. Each sub-sample is further divided into
metallicity intervals chosen to separate regions dominated by individual
components of the disk and halo systems. See text for more details. The
black stripes indicate the mean [C/Fe] for stars in each subset. The
vertical dashed line indicates the level of carbon enhancement used in
this paper to indicate CEMP stars, [C/Fe] $> +0.7$.  Note the relatively
high fraction of CEMP stars among the low-eccentricity stars with
metallicities $-1.8 < $ [Fe/H] $\le -0.8$, which are likely members of the
MWTD. Stars with errors in any of the individual derived components of motion
exceeding 50 \kms\ are excluded. }
\label{cfe_stripe}
\end{figure*}

It is interesting to consider the distribution of carbonicity for stars
in our combined sample in different ranges of orbital eccentricity,
shown in Figure~\ref{cfe_stripe}. It should be recalled that this sample
only includes stars with well-measured kinematics. Note that the
low-eccentricity stars shown in the upper panels of this figure possess
only a few stars that exceed [C/Fe] = $+0.7$ (and hence are considered
CEMP stars), and all but a few of those are found in the metallicity
range $-1.8 < $ [Fe/H] $\le -0.8$ that is expected to apply to the MWTD
population. The relatively large fraction of CEMP stars that may belong
to the MWTD has implications for its formation, but larger samples of
stars and more detailed modeling is required before a definitive
evaluation can be made. The high-eccentricity stars in the lower panels
includes a few stars in this same metallicity range, but most are
probably associated with the inner-halo population. The majority of the
high-eccentricity CEMP stars are found in the metallicity range [Fe/H]
$\le -1.8$ that is expected to include members from both in the inner-
and outer-halo populations.

As noted above, Paper~I made the first published claim that there exists
an increasing frequency of CEMP stars with distance from the Galactic
plane (although most of the weight for this suggestion came from the
addition of stars from the sample of \citealt{beers1992}).
\citet{carollo2012} confirmed and extended this claim using a much
larger sample of stars from SDSS. Here, we examine this question once
again, using our combined sample.

\begin{figure}[!ht]
\epsscale{1.16}
\plotone{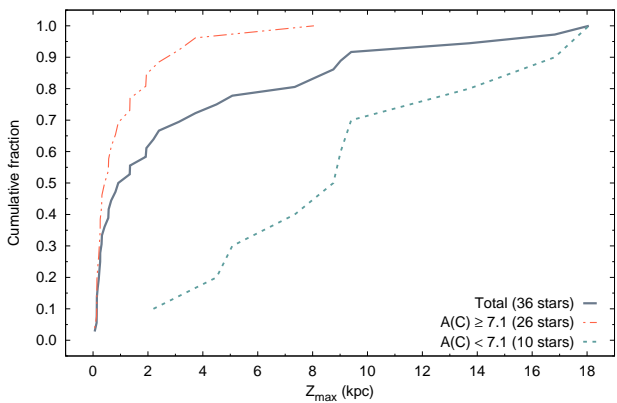}
\caption{Cumulative fractions of CEMP stars, as a function of \zmax,
for stars with [Fe/H] $\le -2.0$ in the combined sample with available measumrements.   
The solid line applies to the full sample, which has a total of 36 CEMP
stars.  The blue-dashed line applies to the 26 CEMP stars with absolute
carbon abundance, A$(C) \ge 7.1$, while the red dashed line applies to
those with A$(C)$ $< 7.1$, a division suggested by \citet{yoon2016} to
distinguish CEMP-$s$ stars from CEMP-no stars.  Note the clear
difference in the behaviors of these two subsets.  See text for more
details.  Stars with errors in any of the individual derived components of motion
exceeding 50 \kms\ are excluded.}
\label{cumul04}
\end{figure}

The solid black line shown in Figure~\ref{cumul04} shows the cumulative
fractions of CEMP stars with [Fe/H] $\le -2.0$ in our combined sample, as a
function of \zmax, which clearly supports the original claim from
Paper~I. This figure also shows lines representing stars from this
sample divided by their absolute carbon abundance, at $A$(C) $\ge 7.1$
(red dot-dashed line) or $A$(C) $< 7.1$ (blue dashed line), the level
suggested by \citet{yoon2016} to effectively separate CEMP-$s$ stars
(those above this value) from the CEMP-no stars (those below this
value). Although the numbers of CEMP stars under consideration is still small
(as indicated in the legend of the figure), there is a rather dramatic
contrast seen between the high-$A$(C) and low-$A$(C) stars, commensurate
with expectation from the study by \citet{carollo2014} that the
inner-halo population of stars comprises a larger fraction of CEMP-$s$
stars than the outer-halo population, which includes greater
relative numbers of CEMP-no stars. A similar exercise applied to the
SDSS sample of CEMP stars, now underway, should prove illuminating.

For convenience, and to inspire future high-resolution spectroscopic
study and radial-velocity monitoring of the relatively bright CEMP stars
we have identified in this work, Table~\ref{tab8} lists the full set of
these stars in our combined sample. Column (1) of this table provides
the star names, column (2) lists their coordinates, and columns (3) and
(4) list the $V$ and $B-V$ colors, respectively. Column (5) lists the
derived metallicity, [Fe/H]$_C$. Columns (6) and (7) list the
carbonicity, [C/Fe]$_C$, and absolute carbon abundances, $A$(C),
respectively. The sub-classification of these CEMP stars, obtained by
application of the \citet{yoon2016} separation of CEMP-$s$ stars from
CEMP-no stars, is listed in column (8). Stars for which a
high-resolution spectrum presently exists (roughly half of the stars,
mostly unpublished, from our group) are indicated in the table. 

Note that a number of {\it known} CEMP stars that are included in our
sample, HE~1327-2326 and HE~1337-0012 (G~64-12), are not included in
this table, since they are sufficiently warm that the carbon enhancement
could not be demonstrated based on the medium-resolution spectroscopy we
have reported on in this paper. There are likely to be others; see the
discussion by \citet{placco2016} of the CEMP status of G~64-12 and
G~64-37. Furthermore, due to the large errors in estimated surface gravities from
our medium-resolution analysis ($\sim 0.5$ dex), we have not explicitly
applied corrections for the depletion of carbon for stars in advanced
evolutionary stages \citep{placco2014}. There are a total of 43 stars in
our combined sample with surface gravity estimates \logg\ $< 2.0$, where
the corrections can become significant.  Of these, seven stars 
(HE~0013-0522, HE~0111-1118, HE~0117-0201, HE~0147-4926, HE~1313-1916,  
HE~2243-0244, and BM-005 = HD~4306) would be considered CEMP stars
if the corrections were applied.  Further attention to these stars is
clearly warranted. 
  
\section{Summary and Discussion}
\label{conclusion}

We have re-analyzed spectra from a previously published sample of 1777
bright metal-poor candidates from the HES \citep{frebel2006}, and
obtained new estimates of their atmospheric parameters \teff, \logg, and
[Fe/H], as well as the carbonicity, [C/Fe]. A large number of stars
(those with [Fe/H] $> -1.0$), whose parameters could not be estimated
previously with the tools in hand, are included in our results. The
carbonicity estimates are refined as well, based on a new grid of
carbon-enhanced synthetic spectra. This sample is combined with stars
from the `weak-metal' candidates of \citet{bidelman1973}, which were
analyzed in a similar fashion by \citet{beers2014}, obtaining a total
sample of 2079 stars. We present a chemodynamical analysis of 1892
stars from this combined sample with suitably precise derived kinematic
properties, and identify new stars that appear to be associated with the
previously suggested halo debris streams from \citet{helmi1999} and
\citet{chiba2000}, as well as with debris stripped from the globular
cluster $\omega$ Cen, discussed by \citet{dinescu2002} and
\citet{majewski2012}. 

It is interesting that a number of the lowest metallicity stars we
identify as part of the $\omega$ Cen debris stream are CEMP-no stars,
which are not expected to form in globular clusters. This may lend credence
to previous speculations that the globular cluster $\omega$ Cen may have
been stripped from a parent dwarf galaxy. If one assumes that this is
the case, this meets with expectations based on the analysis of other
debris streams, such as the Sagittarius Stream, where a number of
authors (e.g., \citealt{deBoer2015}) have suggested that the
low-metallicity stars associated with the parent dwarf were less bound
than the higher-metallicity stars, and stripped early in its interaction
with the Milky Way. Further study of the individual stars which we
suggest may be associated with the putative parent dwarf of $\omega$ Cen
is clearly necessary before this possibility can be confirmed.  

We identify a clear increase in the cumulative frequency of CEMP stars
with declining metallicity, as well as an increase in the fraction of
CEMP stars with distance from the Galactic plane, consistent with
previous results. We also identify a relatively large number of CEMP
stars with kinematics consistent with the MWTD population. This may be
understood if the MWTD were, at least in part, assembled from the debris
of low-mass dwarf galaxies, where CEMP stars (especially CEMP-no stars)
are expected to have formed at high frequency. Although the small number
of stars in this sample precludes stronger conclusions, it will be
interesting to look for this signature in surveys that include larger
samples of likely MWTD stars.

Finally, the 61 CEMP stars in our combined sample are sub-classified
into likely CEMP-$s$ and CEMP-no stars, using the absolute carbon
abundances, $A$(C), as suggested recently by \citet{yoon2016}.

High-resolution spectroscopic analyses of our program stars in the debris
streams that lack this information, as well as those identified as CEMP
stars, are now underway, and will be reported on in due course. Since
these stars are among the brightest examples of the CEMP phenomenon
known, long-term radial-velocity monitoring of these stars, now
underway, should provide valuable information concerning their likely
progenitors.

\acknowledgments 
 
T.C.B., D.C., V.M.P., and S.D. acknowledge partial support for this work
from grants PHY 08-22648; Physics Frontier Center/JINA, and PHY
14-30152; Physics Frontier Center/JINA Center for the Evolution of the
Elements (JINA-CEE), awarded by the US National Science Foundation. S.R.
acknowledges partial support for this work from CAPES, CNPq, FAPESP, and
INCT. Y.S.L. acknowledges partial support from the National Research
Foundation of Korea to the Center for Galaxy Evolution Research and
Basic Science Research Program through the National Research Foundation
of Korea (NRF) funded by the Ministry of Science, ICT \& Future Planning
(NRF-2015R1C1A1A02036658). A.F. acknowledges funding from NSF-CAREER
grant AST 12-55160, and from the Silvermam (1968) Family Career
Development Professorship. Study at RSAA, ANU, of the Galaxy's metal-poor stars
is supported by Australian Research Council grants DP0663562 and DP0984924,  
which J.E.N. is pleased to acknowledge.

\clearpage

\bibliographystyle{apj}




%

\clearpage





\clearpage

\LongTables

\begin{landscape}

\begin{deluxetable}{lllrrrrrrcc}
\tabletypesize{\scriptsize}
\tablecolumns{11}
\tablewidth{0pc}
\tablecaption{Photometric Information and Adopted Reddening \label{tab1}}
\tablehead{
\colhead{Star Name} &
\colhead{Other Name} &
\colhead{HK Survey} &
\colhead{LON} &
\colhead{LAT} &
\colhead{$V$} &
\colhead{$B-V$} &
\colhead{$J$} &
\colhead{$J-K$} &
\colhead{$E(B-V)_S$} &
\colhead{$E(B-V)_A$} \\
\colhead{ } &
\colhead{ } &
\colhead{ } &
\colhead{($^{\circ}$)} &
\colhead{($^{\circ}$)} &
\colhead{ (mag)} &
\colhead{ (mag)} &
\colhead{ (mag)} &
\colhead{ (mag)} &
\colhead{ (mag)} &
\colhead{ (mag)} \\
\colhead{ (1) } &
\colhead{ (2) } &
\colhead{ (3) } &
\colhead{ (4) } &
\colhead{ (5) } &
\colhead{ (6) } &
\colhead{ (7) } &
\colhead{ (8) } &
\colhead{ (9) } &
\colhead{ (10) } &
\colhead{ (11) } 
 }
\startdata
HE~0000-3017  & CD-30:19801             &                            &   14.8 & $-$79.1 & 10.36 &  0.40 &  9.546 &  0.275 &  0.015 & 0.02\\
HE~0000-4401  & CD-44:15435             &                            &  330.2 & $-$70.7 & 10.66 &  0.55 &  9.717 &  0.302 &  0.010 & 0.01\\
HE~0000-5703  & HD~225032               &                            &  315.9 & $-$59.1 &  9.40 &  0.24 &  9.001 &  0.109 &  0.010 & 0.01\\
HE~0001-4157  & CD-42:16578             &                            &  333.7 & $-$72.5 & 10.52 &  0.67 &  9.385 &  0.376 &  0.012 & 0.01\\
HE~0001-4449  & CD-45:15231             &                            &  328.4 & $-$70.2 & 11.14 &  0.49 & 10.192 &  0.296 &  0.012 & 0.01\\
HE~0001-5640  & CD-57:8943              &                            &  315.9 & $-$59.5 & 10.50 &  0.41 &  9.571 &  0.285 &  0.011 & 0.01\\
HE~0002-3233  &                         & CS~22961-023               &    2.9 & $-$78.8 & 12.00 &  0.41 & 11.068 &  0.323 &  0.016 & 0.02\\
HE~0002-3822  & CD-38:15729             &                            &  341.8 & $-$75.3 & 10.79 &  0.53 &  9.808 &  0.335 &  0.016 & 0.02\\
HE~0002-5625  &                         &                            &  315.9 & $-$59.8 & 12.60 &  0.69 & 11.425 &  0.424 &  0.010 & 0.01\\
HE~0003-0503  & BD-05:6112              &                            &   95.0 & $-$65.2 & 10.70 &  0.74 &  8.990 &  0.602 &  0.030 & 0.03\\
\enddata
\end{deluxetable}

\pagestyle{empty}
\voffset=-25pt

\begin{deluxetable}{@{}l@{}l@{}rrrrrrrrrrrrrrrrrrl@{}}
\tabletypesize{\tiny}
\tablecolumns{21}
\tablewidth{0pc}
\tablecaption{Radial Velocities, Line Indices, Atmospheric Parameters, and Type Assignments \label{tab2}}
\tablehead{
\colhead{Star Name} &
\colhead{Note} &
\colhead{RV$_{M}$} &
\colhead{RV$_{R}$} &
\colhead{RV$_{H}$} &
\colhead{ KP } &
\colhead{ HP2 } &
\colhead{ GP } &
\colhead{ Teff$_S$} &
\colhead{ logg$_S$} &
\colhead{ [Fe/H]$_S$ } &
\colhead{ Teff$_R$} &
\colhead{ logg$_R$} &
\colhead{ [Fe/H]$_R$ } &
\colhead{ Teff$_H$} &
\colhead{ logg$_H$} &
\colhead{ [Fe/H]$_H$ } &
\colhead{ Teff$_C$} &
\colhead{ logg$_C$} &
\colhead{ [Fe/H]$_C$ } &
\colhead{ TYPE} \\
\colhead{ } &
\colhead{ } &
\colhead{(km~s$^{-1}$) } &
\colhead{(km~s$^{-1}$) } &
\colhead{(km~s$^{-1}$) } &
\colhead{ ({\AA}) } &
\colhead{ ({\AA}) } &
\colhead{ ({\AA}) } &
\colhead{ (K)}&
\colhead{ (cgs)} &
\colhead{  } &
\colhead{ (K)} &
\colhead{ (cgs)} &
\colhead{  } &
\colhead{ (K)} &
\colhead{ (cgs)} &
\colhead{  } &
\colhead{ (K)} &
\colhead{ (cgs)} &
\colhead{ }  &
\colhead{  } \\
\colhead{(1)} &
\colhead{(2)} &
\colhead{(3)} &
\colhead{(4)} &
\colhead{(5)} &
\colhead{(6)} &
\colhead{(7)} &
\colhead{(8)} &
\colhead{(9)} &
\colhead{(10)} &
\colhead{(11)} &
\colhead{(12)} &
\colhead{(13)} &
\colhead{(14)} &
\colhead{(15)} &
\colhead{(16)} &
\colhead{(17)} &
\colhead{(18)} &
\colhead{(19)} &
\colhead{(20)} &
\colhead{(21)} 
}
\startdata 
HE~0000-3017  &           & $  21.8  $&$  \dots  $&$  \dots  $&  6.57 &  4.99 &  1.89 &  6666 &  3.99  &$  -0.18   $& \dots   &  \dots   &$   \dots   $&   6875  &  4.60  &$  +0.23  $&  6776 &  4.41  &$   +0.20  $& D      \\
HE~0000-4401  &           & $  -1.0  $&$   -2.1  $&$  \dots  $&  8.29 &  3.60 &  3.55 &  6199 &  3.77  &$  -0.13   $&  6168   &   4.12   &$   -0.05   $&  \dots  & \dots  &$  \dots  $&  6227 &  4.14  &$   +0.26  $& D      \\
HE~0000-5703  &           & $  30.3  $&$   26.4  $&$  \dots  $&  2.59 &  9.37 &  1.16 &  7758 &  4.08  &$  -0.03   $&  7474   &   4.25   &$   -0.10   $&  \dots  & \dots  &$  \dots  $&  8060 &  4.53  &$   +0.30  $& D      \\
HE~0001-4157  &           & $   2.0  $&$  -10.8  $&$  \dots  $&  9.15 &  1.73 &  5.47 &  5725 &  3.81  &$  -0.10   $&  5710   &   3.99   &$   +0.10   $&  \dots  & \dots  &$  \dots  $&  5670 &  4.19  &$   +0.30  $& D      \\
HE~0001-4449  &           & $   5.0  $&$   -9.4  $&$  \dots  $&  7.53 &  3.53 &  2.92 &  6227 &  3.65  &$  -0.58   $&  5959   &   3.66   &$   -0.62   $&  \dots  & \dots  &$  \dots  $&  6260 &  3.99  &$   -0.28  $& TO     \\
HE~0001-5640  &           & $ -14.0  $&$  [23.3] $&$  \dots  $&  7.80 &  4.06 &  2.82 &  6377 &  3.86  &$  -0.26   $&  6176   &   3.65   &$   +0.14   $&  \dots  & \dots  &$  \dots  $&  6436 &  4.25  &$   +0.11  $& D      \\
HE~0002-3233  &           & $  61.1  $&$  \dots  $&$  \dots  $&  1.30 &  4.24 &  0.48 &  6349 &  3.65  &$  -2.54   $& \dots   &  \dots   &$   \dots   $&  \dots  & \dots  &$  \dots  $&  6404 &  4.00  &$   -2.70  $& TO     \\
HE~0002-3822  &           & $  -8.0  $&$  \dots  $&$  \dots  $&  7.80 &  3.39 &  3.20 &  6132 &  3.87  &$  -0.59   $& \dots   &  \dots   &$   \dots   $&  \dots  & \dots  &$  \dots  $&  6148 &  4.26  &$   -0.30  $& D      \\
HE~0002-5625  &           & $  22.4  $&$  \dots  $&$  \dots  $&  9.25 &  1.65 &  5.55 &  5750 &  4.02  &$  -0.14   $& \dots   &  \dots   &$   \dots   $&  \dots  & \dots  &$  \dots  $&  5699 &  4.45  &$   +0.25  $& D      \\
HE~0003-0503  &           & $  30.6  $&$   34.4  $&$  \dots  $&  6.82 &  3.39 &  4.21 &  5972 &  2.63  &$  -1.10   $& (4793)  &  (3.46)  &$  (-0.19)  $&  \dots  & \dots  &$  \dots  $&  5960 &  2.72  &$   -0.92  $& G      \\
\enddata                                      
\tablecomments{Parentheses around a listed quantity indicate that it is regarded with some suspicion, while
brackets indicate that it is considered as possibly flawed.  See text for more details.}

\end{deluxetable}

\begin{deluxetable}{lccccrcccc}
\tabletypesize{\footnotesize}
\tablecolumns{10}
\tablewidth{0pc}
\tablecaption{Carbon Abundance Ratios and Absolute Carbon Abundance Estimates \label{tab3}}
\tablehead{
\colhead{Star Name} &
\colhead{[Fe/H]$_S$} &
\colhead{[C/Fe]$_S$} &
\colhead{DETECT} &
\colhead{CC} &
\colhead{EQW} &
\colhead{[Fe/H]$_C$} &
\colhead{[C/Fe]$_C$} &
\colhead{A(C)} &
\colhead{CEMP} \\
\colhead{ (1) } &
\colhead{ (2) } &
\colhead{ (3) } &
\colhead{ (4) } &
\colhead{ (5) } &
\colhead{ (6) } &
\colhead{ (7) } &
\colhead{ (8) } &
\colhead{ (9) } &
\colhead{ (10)  } 
}
\startdata
HE~0000-3017  &$  -0.18  $&$ +0.15   $&  D     &$   0.995 $&   2.33   &$  +0.20   $&$  -0.12   $&  8.52  &  N   \\
HE~0000-4401  &$  -0.13  $&$ +0.08   $&  D     &$   0.996 $&   4.14   &$  +0.26   $&$  -0.19   $&  8.50  &  N   \\
HE~0000-5703  &$  -0.03  $&$   \dots $&  X     &$   \dots $&    \dots &$  +0.30   $&$    \dots $&  \dots &  X   \\
HE~0001-4157  &$  -0.10  $&$ +0.04   $&  D     &$   0.999 $&   6.42   &$  +0.30   $&$  -0.23   $&  8.50  &  N   \\
HE~0001-4449  &$  -0.58  $&$ +0.25   $&  D     &$   0.994 $&   3.26   &$  -0.28   $&$ \phantom{-}0.00 $&  8.15  &  N   \\
HE~0001-5640  &$  -0.26  $&$ +0.19   $&  D     &$   0.997 $&   3.48   &$  +0.11   $&$  -0.07   $&  8.46  &  N   \\
HE~0002-3233  &$  -2.54  $&$ +1.29   $&  D:    &$   0.866 $&   0.76   &$  -2.70   $&$  +1.11   $&  6.84  &  U   \\
HE~0002-3822  &$  -0.59  $&$ +0.17   $&  D     &$   0.994 $&   3.66   &$  -0.30   $&$  -0.09   $&  8.04  &  N   \\
HE~0002-5625  &$  -0.14  $&$ -0.03   $&  D     &$   0.999 $&   6.71   &$  +0.25   $&$  -0.30   $&  8.38  &  N   \\
HE~0003-0503  &$  -1.10  $&$ +1.42   $&  D     &$   0.971 $&   5.20   &$  -0.92   $&$  +1.25   $&  8.75  &  C   \\
\enddata                                                                                
\tablecomments{A ``:'' following the DETECT code indicates that either the CC or EQW parameters do not meet the minimum required 
value for confident detection.  See text for more details.}
\end{deluxetable}

\begin{deluxetable}{llrrccccc}
\tabletypesize{\scriptsize}
\tablecolumns{9}
\tablewidth{0pc}
\tablecaption{Parallaxes and Distance Estimates for Stars with Hipparcos Measurements \label{tab4}}
\tablehead{
\colhead{Star Name} &
\colhead{Type} &
\colhead{$\pi_{\rm HIP}$} &
\colhead{$\sigma_{\pi_{\rm HIP}}$ } &
\colhead{$\sigma_{\pi_{\rm HIP}}/{\pi_{\rm HIP}}$} &
\colhead{$D_{\rm HIP}$ }&
\colhead{$\sigma_{{\rm D}_{\rm HIP}}$ } &
\colhead{$D_{\rm pho}$ } &
\colhead{$\sigma_{{\rm D}_{\rm pho}}$ } \\
\colhead{ } &
\colhead{ } &
\colhead{(mas)} &
\colhead{(mas)} &
\colhead{ } &
\colhead{(kpc)} &
\colhead{(kpc)} &
\colhead{(kpc)} &
\colhead{(kpc)} \\
\colhead{(1) } &
\colhead{(2) } &
\colhead{(3) } &
\colhead{(4) } &
\colhead{(5) } &
\colhead{(6) } &
\colhead{(7) } &
\colhead{(8) } &
\colhead{(9) } 
}
\startdata
HE~0035-0834        & D     &    9.51 &    1.11 &    0.12 &   0.105 &   0.012 &   0.084 &  0.013 \\ 
HE~0115-5135        & G     &    6.67 &    1.30 &    0.19 &   0.150 &   0.029 &   0.285 &  0.043 \\ 
HE~0134-2142        & D     &    7.57 &    1.55 &    0.20 &   0.132 &   0.027 &   0.150 &  0.023 \\ 
HE~0246-5114        & D     &    4.31 &    0.66 &    0.15 &   0.232 &   0.036 &   0.199 &  0.030 \\ 
HE~0422-4205        & D     &    5.61 &    1.13 &    0.20 &   0.178 &   0.036 &   0.158 &  0.024 \\ 
HE~0429-4149        & D     &   13.80 &    0.95 &    0.07 &   0.072 &   0.005 &   0.064 &  0.010 \\ 
HE~0435-4121        & D     &    5.33 &    1.06 &    0.20 &   0.188 &   0.037 &   0.181 &  0.027 \\ 
HE~0455-3157        & D     &   10.05 &    1.31 &    0.13 &   0.100 &   0.013 &   0.076 &  0.011 \\ 
HE~0457-3209        & D     &    6.42 &    1.18 &    0.18 &   0.156 &   0.029 &   0.134 &  0.020 \\ 
HE~0511-4835        & D     &   11.23 &    0.92 &    0.08 &   0.089 &   0.007 &   0.099 &  0.015 \\ 
HE~0520-5617        & D     &    4.55 &    0.50 &    0.11 &   0.220 &   0.024 &   0.142 &  0.021 \\ 
HE~1108-3217        & D     &    6.80 &    0.60 &    0.09 &   0.147 &   0.013 &   0.049 &  0.007 \\ 
HE~1120-0858        & D     &    8.32 &    1.49 &    0.18 &   0.120 &   0.022 &   0.177 &  0.027 \\ 
HE~1211-3038        & D     &   13.81 &    1.50 &    0.11 &   0.072 &   0.008 &   0.100 &  0.015 \\ 
HE~1223-0133        & TO    &    9.05 &    1.11 &    0.12 &   0.110 &   0.014 &   0.125 &  0.019 \\ 
HE~1349-1827        & FHB   &    4.57 &    0.76 &    0.17 &   0.219 &   0.036 &   \dots &  \dots \\ 
HE~1411-0542        & TO    &   23.33 &    0.53 &    0.02 &   0.043 &   0.001 &   0.022 &  0.003 \\ 
HE~1450-1808        & D     &   25.31 &    1.85 &    0.07 &   0.040 &   0.003 &   0.027 &  0.004 \\ 
HE~2231-0149        & D     &    9.38 &    0.47 &    0.05 &   0.107 &   0.005 &   0.283 &  0.042 \\ 
HE~2255-1758        & D     &    5.14 &    0.88 &    0.17 &   0.195 &   0.033 &   0.175 &  0.026 \\ 
HE~2307-4543        & D     &   22.13 &    1.31 &    0.06 &   0.045 &   0.003 &   0.046 &  0.007 \\ 
HE~2327-4203        & D     &   10.17 &    1.19 &    0.12 &   0.098 &   0.012 &   0.145 &  0.022 \\ 
HE~2332-4431        & D     &   10.06 &    1.78 &    0.18 &   0.099 &   0.018 &   0.127 &  0.019 \\ 
HE~2333-4047        & D     &    8.67 &    1.65 &    0.19 &   0.115 &   0.022 &   0.110 &  0.017 \\ 
HE~2333-4325        & TO    &    6.47 &    1.31 &    0.20 &   0.155 &   0.031 &   0.149 &  0.022 \\ 
\enddata

\end{deluxetable}

\begin{deluxetable}{llccrrccc}
\tabletypesize{\scriptsize}
\tablecolumns{9}
\tablewidth{0pc}
\tablecaption{Distance Estimates and Proper Motions \label{tab5}}
\tablehead{
\colhead{Star Name} &
\colhead{Type} &
\colhead{$D_{\rm pho}$ }&
\colhead{$\sigma_{{\rm D}_{\rm pho}}$ } &
\colhead{$\mu_{\alpha}$ } &
\colhead{$\mu_{\delta}$ } &
\colhead{$\sigma_{\mu_{\alpha}}$ } &
\colhead{$\sigma_{\mu_{\delta}}$ } &
\colhead{PM Source} \\
\colhead{ } &
\colhead{ } &
\colhead{(kpc)} &
\colhead{(kpc)} &
\colhead{(mas yr$^{-1}$)} &
\colhead{(mas yr$^{-1}$)} &
\colhead{(mas yr$^{-1}$)} &
\colhead{(mas yr$^{-1}$)} &
\colhead{ } \\ 
\colhead{(1) } &
\colhead{(2) } &
\colhead{(3) } &
\colhead{(4) } &
\colhead{(5) } &
\colhead{(6) } &
\colhead{(7) } &
\colhead{(8) } &
\colhead{(9) } 
}
\startdata
HE~0000-3017  &  D     &   0.249 &   0.037 &$  20.0 $&$     1.3 $&     1.1 &     0.8 &  U    \\
HE~0000-4401  &  D     &   0.174 &   0.026 &$  23.6 $&$    11.5 $&     1.0 &     2.9 &  U    \\
HE~0000-5703  &  D     &   0.253 &   0.038 &$  13.9 $&$     2.6 $&     0.9 &     1.0 &  U    \\
HE~0001-4157  &  D     &   0.107 &   0.016 &$ -19.9 $&$   -54.8 $&     0.8 &     0.8 &  U    \\
HE~0001-4449  &  TO    &   0.223 &   0.033 &$  -1.0 $&$     0.7 $&     1.1 &     1.0 &  U    \\
HE~0001-5640  &  D     &   0.249 &   0.037 &$  40.2 $&$   -13.0 $&     1.0 &     1.2 &  U    \\
HE~0002-3233  &  TO    &   0.430 &   0.065 &$  57.4 $&$   -31.6 $&     1.3 &     1.5 &  U    \\
HE~0002-3822  &  D     &   0.158 &   0.024 &$ -36.2 $&$    -8.9 $&     1.0 &     1.5 &  U    \\
HE~0002-5625  &  D     &   0.258 &   0.039 &$  11.8 $&$    -4.9 $&     1.4 &     1.4 &  U    \\
HE~0003-0503  &  G     &   0.163 &   0.024 &$   9.5 $&$    -1.5 $&     1.8 &     1.3 &  U    \\
\enddata                                                                              

\tablecomments{Sources of proper motions:  U = UCAC4, S = SPM4, H = Hipparcos or Tycho II}
\end{deluxetable}

\pagestyle{empty}

\begin{deluxetable}{lrrrrrrrrrrrrrrc}
\tablewidth{0pt} 
\tabletypesize{\scriptsize} 
\tablecolumns{16} 
\tablecaption{Space Motions and Orbital Parameters \label{tab6}}
\tablehead{
\colhead{Star Name} &
\colhead{$R$ } &
\colhead{$Z$ } &
\colhead{$U$ } &
\colhead{$V$ } &
\colhead{$W$ } &
\colhead{$\sigma(U)$ } &
\colhead{$\sigma(V)$ } &
\colhead{$\sigma(W)$ } &
\colhead{$V_{\rm R}$} &
\colhead{$V_{\phi}$} &
\colhead{$r_{\rm peri}$} &
\colhead{$r_{\rm apo}$} &
\colhead{$Z_{\rm max}$} &
\colhead{$e$} &
\colhead{INOUT} \\ 
\colhead{ } &
\colhead{(kpc)} &
\colhead{(kpc)} &
\colhead{(km~s$^{-1}$)} &
\colhead{(km~s$^{-1}$)} &
\colhead{(km~s$^{-1}$)} &
\colhead{(km~s$^{-1}$)} &
\colhead{(km~s$^{-1}$)} &
\colhead{(km~s$^{-1}$)} &
\colhead{(km~s$^{-1}$)} &
\colhead{(km~s$^{-1}$)} &
\colhead{(kpc)} &
\colhead{(kpc)} &
\colhead{(kpc)} &
\colhead{ }     &
\colhead{ } \\
\colhead{(1)} &
\colhead{(2)} &
\colhead{(3)} &
\colhead{(4)} &
\colhead{(5)} &
\colhead{(6)} &
\colhead{(7)} &
\colhead{(8)} &
\colhead{(9)} &
\colhead{(10)} &
\colhead{(11)} &
\colhead{(12)} &
\colhead{(13)} &
\colhead{(14)} &  
\colhead{(15)} &  
\colhead{(16)}   
}
\startdata
HE~0000-3017  &    8.454 &$  -0.244 $&$     8 $&$     4 $&$   -19 $&     4 &     2 &    10 &$     9 $&$   224 $&    8.30 &    8.82 &    0.36 &    0.03 &  1 \\ 
HE~0000-4401  &    8.450 &$  -0.164 $&$    12 $&$    12 $&$     3 $&     4 &     2 &     5 &$    12 $&$   232 $&    8.34 &    9.32 &    0.18 &    0.06 &  1 \\ 
HE~0000-5703  &    8.407 &$  -0.217 $&$    -3 $&$    -2 $&$   -20 $&     3 &     2 &     4 &$    -6 $&$   218 $&    8.10 &    8.47 &    0.34 &    0.02 &  1 \\ 
HE~0001-4157  &    8.471 &$  -0.102 $&$   -26 $&$    -6 $&$    26 $&     3 &     3 &     5 &$   -27 $&$   213 $&    7.50 &    8.99 &    0.35 &    0.09 &  1 \\ 
HE~0001-4449  &    8.436 &$  -0.210 $&$    -7 $&$    15 $&$    16 $&     2 &     1 &     5 &$    -8 $&$   235 $&    8.39 &    9.51 &    0.32 &    0.06 &  1 \\ 
HE~0001-5640  &    8.410 &$  -0.215 $&$    32 $&$   -17 $&$    18 $&     7 &     6 &     9 &$    30 $&$   203 $&    6.90 &    8.81 &    0.33 &    0.12 &  1 \\ 
HE~0002-3233  &    8.417 &$  -0.422 $&$    52 $&$   -98 $&$   -68 $&    12 &    17 &    10 &$    52 $&$   122 $&    3.54 &    8.84 &    1.61 &    0.43 &  1 \\ 
HE~0002-3822  &    8.462 &$  -0.153 $&$   -34 $&$    19 $&$    21 $&     5 &     2 &    10 &$   -34 $&$   239 $&    8.04 &   10.45 &    0.35 &    0.13 &  1 \\ 
HE~0002-5625  &    8.407 &$  -0.223 $&$    -6 $&$    -7 $&$   -12 $&     4 &     4 &     9 &$    -9 $&$   213 $&    7.72 &    8.48 &    0.27 &    0.05 &  1 \\ 
HE~0003-0503  &    8.506 &$  -0.148 $&$    -2 $&$    22 $&$   -26 $&     2 &     2 &     5 &$     0 $&$   242 $&    8.51 &   10.19 &    0.43 &    0.09 &  1 \\ 
\enddata
\tablecomments{INOUT takes on a value of ``1'' if the star is accepted for the kinematic analysis, ``0'' if not.}  

\end{deluxetable}

\begin{deluxetable}{lcrcccccccc}
\tabletypesize{\scriptsize}
\tablecolumns{11}
\tablewidth{0pt}
\tablecaption{Parameters for Stars in the Identified Streams \label{tab7}}
\tablehead{
\colhead{Star Name} &
\colhead{RA (2000) DEC} &
\colhead{$V$} &
\colhead{$B-V$} &
\colhead{[Fe/H]$_C$} &
\colhead{[C/Fe]$_C$} &
\colhead{A$(C)$} &
\colhead{Class} &
\colhead{L${_\bot}$} &
\colhead{L${_Z}$} &
\colhead{E} \\ 
\colhead{ } &
\colhead{ } &
\colhead{(mag)} &
\colhead{(mag)} &
\colhead{ } &
\colhead{ } &
\colhead{ } &
\colhead{ } &
\colhead{(kpc km s$^{-1}$)} &
\colhead{~~~(kpc km s$^{-1}$)} &
\colhead{(10$^5$ km$^{2}$ s$^{-2}$)} \\
\colhead{(1) } &
\colhead{(2) } &
\colhead{(3) } &
\colhead{(4) } &
\colhead{(5) } &
\colhead{(6) } &
\colhead{(7) } &
\colhead{(8) } &
\colhead{(9) } &
\colhead{~~~(10)} &
\colhead{(11)}  
}
\startdata
\multicolumn{11}{c}{Helmi et al. Debris Stream}\\
\hline
HE~0012-5643\tablenotemark{a}       & 00 15 17.1 $-$56 26 27 & 12.29 & 0.46 & $-$2.97 & $+$1.41 & 6.87  & CEMP-$s$  &  2466   &  1132    & $-$0.81 \\ 
HE~0017-3646                        & 00 20 26.1 $-$36 30 20 & 13.02 & 0.54 & $-$2.48 & $-$0.54 & 5.41  & \dots     &  2729   &  1169    & $-$0.92 \\ 
HE~0048-1109\tablenotemark{a}       & 00 51 26.4 $-$10 53 14 & 10.83 & 0.49 & $-$1.97 & $-$0.10 & 6.36  & \dots     &  2187   &  1337    & $-$1.07 \\ 
BM-028                              & 02 47 37.4 $-$36 06 27 &  9.94 & 0.46 & $-$1.58 & $+$0.24 & 7.31  & \dots     &  2259   &  1020    & $-$1.11 \\
HE~0324-0122                        & 03 27 02.3 $+$01 32 33 & 12.13 & 0.72 & $-$2.11 & $+$0.37 & 6.69  & \dots     &  2311   &  1493    & $-$1.06 \\ 
BM-209                              & 14 36 48.5 $-$29 06 47 &  8.02 & 0.64 & $-$1.91 & $+$0.06 & 6.66  & \dots     &  2152   &  1146    & $-$1.14 \\ 
HE~2215-3842                        & 22 18 20.9 $-$38 27 55 & 13.40 & 0.68 & $-$2.24 & $+$0.45 & 6.64  & \dots     &  2062   &  1093    & $-$0.97 \\ 
BM-308                              & 22 37 08.1 $-$40 30 39 &  9.11 & 0.79 & $-$2.12 & $-$0.10 & 6.39  & \dots     &  2313   &  1124    & $-$0.73 \\ 
\hline
\multicolumn{11}{c}{Helmi et al. Trail}\\
\hline
HE~0033-2141\tablenotemark{a}       & 00 35 42.1 $-$21 24 58 & 12.29 & 0.72 & $-$2.73 & $+$0.15 & 5.85  & \dots     &  1407   &  1559    & $-$1.11 \\ 
HE~0050-0918                        & 00 52 41.7 $-$09 02 23 & 11.06 & 0.71 & $-$2.08 & $-$0.26 & 6.09  & \dots     &  1653   &  1122    & $-$1.14 \\ 
HE~1210-2729\tablenotemark{a}       & 12 13 07.9 $-$27 45 50 & 12.54 & 0.86 & $-$2.95 & $-$0.18 & 5.31  & \dots     &  1589   &  1172    & $-$1.09 \\ 
BM-235                              & 17 52 35.9 $-$69 01 45 &  9.48 & 1.05 & $-$1.83 & $-$0.44 & 6.39  & \dots     &  1925   &  1130    & $-$1.17 \\ 
HE~2234-4757\tablenotemark{a}       & 22 37 20.4 $-$47 41 38 & 12.39 & 0.92 & $-$2.59 & $-$0.26 & 5.57  & \dots     &  1455   &  1441    & $-$1.10 \\ 
\hline
\multicolumn{11}{c}{$\omega$ Cen Debris Stream}\\
\hline
HE~0007-1752\tablenotemark{a}       & 00 10 17.6 $-$17 35 38 & 11.54 & 0.65 & $-$2.47 & $+$0.54 & 6.50  & \dots     &  2333   &   \phn\phn33     & $-$1.17 \\ 
HE~0039-0216\tablenotemark{a}       & 00 41 53.6 $-$02 00 33 & 13.35 & 0.37 & $-$2.62 & $+$1.40 & 7.21  & CEMP-$s$  &  2993   &   \phn230    & $-$0.93 \\ 
HE~0429-4620                        & 04 30 48.6 $-$46 13 53 & 13.10 & 0.62 & $-$2.42 & $+$0.58 & 6.59  & \dots     &  1912   &   $-$236 & $-$1.14 \\ 
HE~1120-0153\tablenotemark{a}       & 04 38 55.7 $-$13 20 48 & 11.68 & 0.44 & $-$2.39 & $-$0.17 & 5.89  & \dots     &  2023   &   \phn140    & $-$0.97 \\ 
BM-056                              & 05 10 49.6 $-$37 49 03 &  9.50 & 0.86 & $-$2.00 & $-$0.16 & 7.62  & \dots     &  2377   &   $-$102 & $-$1.12 \\
BM-121\tablenotemark{a}             & 09 53 39.2 $-$22 50 08 &  9.39 & 1.16 & $-$2.69 & $-$0.53 & 5.65  & \dots     &  1550   &   \phn$-$84  & $-$1.37 \\
HE~1120-0153\tablenotemark{a}       & 11 22 43.2 $-$02 09 36 & 11.68 & 0.44 & $-$2.88 & $+$1.09 & 6.64  & CEMP-no   &  1569   &   \phn$-$41  & $-$1.12 \\ 
HE~1401-0010\tablenotemark{a}       & 14 04 03.4 $-$00 24 25 & 13.51 & 0.41 & $-$2.44 & $+$0.73 & 6.72  & CEMP-no   &  2681   &   \phn138    & $-$0.79 \\ 
HE~2138-0314\tablenotemark{a}       & 21 40 41.5 $-$03 01 17 & 13.23 & 0.57 & $-$3.07 & $+$0.90 & 6.27  & CEMP-no   &  2138   &   \phn$-$70  & $-$1.13 \\ 
HE~2315-4306                        & 23 18 19.0 $-$42 50 27 & 11.28 & 0.65 & $-$2.36 & $+$0.37 & 6.43  & \dots     &  1712   &   $-$202 & $-$1.33 \\ 
HE~2319-5228\tablenotemark{a}       & 23 21 58.1 $-$52 11 43 & 13.25 & 0.90 & $-$3.39 & $+$1.47 & 6.51  & CEMP-no   &  2170   &   \phn114    & $-$1.07 \\ 
HE~2322-6125\tablenotemark{a}       & 23 25 34.6 $-$61 09 10 & 12.47 & 0.63 & $-$2.50 & $+$0.17 & 6.10  & \dots     &  2145   &   \phn$-$80  & $-$1.20 \\ 
\enddata

\tablenotetext{a}{A high-resolution spectrum exists for this star.}
\end{deluxetable}


\begin{deluxetable}{lcrccccl}
\tabletypesize{\scriptsize}
\tablecolumns{8}
\tablewidth{0pc}
\tablecaption{CEMP Stars and Sub-classifications for Stars in the Combined Sample \label{tab8}}
\tablehead{
\colhead{Star Name} &
\colhead{RA (2000) DEC} &
\colhead{$V$} &
\colhead{$B-V$} &
\colhead{[Fe/H]$_C$} &
\colhead{[C/Fe]$_C$} &
\colhead{A$(C)$} &
\colhead{Class} \\
\colhead{(1) } &
\colhead{(2) } &
\colhead{(3) } &
\colhead{(4) } &
\colhead{(5) } &
\colhead{(6) } &
\colhead{(7) } &
\colhead{(8) } 
}
\startdata
HE~0013-0257\tablenotemark{a}  &  00 16 04.2 $-$02 41 05   & 12.71   &  0.79     &  $-3.42  $&  0.75  &  5.76  &  CEMP-no \\
HE~0015-0048\tablenotemark{a}  &  00 18 01.4 $+$01 05 08   & 13.20   &  0.76     &  $-2.70  $&  0.81  &  6.54  &  CEMP-no \\  
HE~0027-1221\tablenotemark{a}  &  00 30 31.1 $-$12 05 11   & 13.03   &  0.67     &  $-2.50  $&  2.55  &  8.47  &  CEMP-$s$\\
HE~0030-5441                   &  00 33 20.1 $-$54 24 43   & 10.62   &  0.36     &  $-1.26  $&  0.77  &  7.94  &  CEMP-$s$\\
HE~0038-0345                   &  00 41 09.3 $-$03 29 00   & 11.42   &  0.78     &  $-2.66  $&  0.76  &  6.53  &  CEMP-no \\
HE~0039-2635\tablenotemark{a}  &  00 41 39.8 $-$26 18 54   & 12.18   &  1.15     &  $-3.94  $&  3.15  &  7.64  &  CEMP-$s$\\
HE~0054-2542\tablenotemark{a}  &  00 57 18.1 $-$25 26 10   & 12.63   &  0.95     &  $-3.52  $&  2.63  &  7.54  &  CEMP-$s$\\
HE~0058-3449                   &  01 01 21.7 $-$34 33 11   & 13.21   &  0.71     &  $-2.22  $&  0.96  &  7.17  &  CEMP-$s$\\
HE~0228-0149                   &  02 30 56.0 $-$01 36 03   & 12.68   &  0.51     &  $-1.79  $&  0.87  &  7.52  &  CEMP-$s$\\
BM-024\tablenotemark{a}        &  02 39 02.5 $-$49 27 46   & 10.11   &  0.77     &  $-2.52  $&  1.71  &  7.62  &  CEMP-$s$\\
                               &                           &         &           &           &        &        &          \\
HE~0247-0254                   &  02 50 16.9 $-$02 41 50   & 13.38   &  0.60     &  $-1.49  $&  1.51  &  8.45  &  CEMP-$s$\\
BM-043\tablenotemark{a}        &  04 13 13.1 $+$06 36 02   &  9.10   &  1.29     &  $-2.46  $&  2.43  &  8.40  &  CEMP-$s$\\
HE~0412-0138                   &  04 15 33.5 $+$01 45 58   & 10.50   &  0.75     &  $-1.52  $&  0.78  &  7.69  &  CEMP-$s$\\
HE~0414-0343\tablenotemark{a}  &  04 17 16.5 $-$03 36 31   & 10.63   &  1.09     &  $-3.23  $&  1.75  &  6.95  &  CEMP-$s$\\
HE~0420-0123\tablenotemark{a}  &  04 23 14.5 $+$01 30 48   & 11.35   &  0.79     &  $-2.75  $&  2.47  &  8.15  &  CEMP-$s$\\
HE~0440-3426\tablenotemark{a}  &  04 42 08.2 $-$34 21 14   & 11.42   &  1.18     &  $-2.46  $&  1.12  &  7.09  &  CEMP-no \\
HE~0448-4806\tablenotemark{a}  &  04 49 33.1 $-$48 01 08   & 12.78   &  0.62     &  $-2.76  $&  2.88  &  8.55  &  CEMP-$s$\\
HE~0543-5350                   &  05 44 42.0 $-$53 49 01   & 11.97   &  0.48     &  $-2.39  $&  0.88  &  6.92  &  CEMP-$s$\\
BM-074                         &  06 04 07.1 $-$20 37 14   &  8.69   &  0.48     &  $-1.01  $&  1.01  &  8.43  &  CEMP-$s$\\
BM-083                         &  06 34 55.5 $-$45 18 30   &  7.19   &  0.81     &  $-2.12  $&  0.94  &  7.25  &  CEMP-$s$\\  
                               &                           &         &           &           &        &        &          \\
BM-091                         &  07 34 28.9 $-$13 52 13   &  6.70   &  0.47     &  $-1.24  $&  1.09  &  8.28  &  CEMP-$s$\\
HE~0900-0001                   &  09 02 41.3 $-$00 13 35   & 12.70   &  0.39     &  $-1.64  $&  1.09  &  7.88  &  CEMP-$s$\\
BM-107                         &  09 23 02.1 $-$49 03 31   &  8.89   &  0.33     &  $-1.01  $&  0.88  &  8.30  &  CEMP-$s$\\
HE~0920-0506                   &  09 23 06.0 $-$05 19 33   & 11.50   &  0.68     &  $-1.39  $&  1.19  &  8.23  &  CEMP-$s$\\
HE~1109-0025                   &  11 12 06.7 $-$00 41 30   & 10.64   &  0.45     &  $-1.25  $&  1.06  &  8.24  &  CEMP-$s$\\
HE~1114-2757                   &  11 17 00.8 $-$28 14 12   & 10.47   &  0.62     &  $-1.27  $&  0.85  &  8.00  &  CEMP-$s$\\
HE~1119-0218                   &  11 22 27.0 $+$02 02 10   & 11.39   &  0.50     &  $-1.44  $&  1.22  &  8.21  &  CEMP-$s$\\
HE~1143-0114\tablenotemark{a}  &  11 46 31.7 $+$00 57 30   & 12.42   &  0.53     &  $-2.44  $&  2.50  &  8.50  &  CEMP-$s$\\
HE~1154-2951\tablenotemark{a}  &  11 56 39.4 $-$30 08 31   & 10.49   &  0.45     &  $-2.54  $&  2.00  &  7.90  &  CEMP-$s$\\
HE~1225-0155\tablenotemark{a}  &  12 28 04.8 $+$01 38 33   & 12.95   &  0.74     &  $-2.68  $&  0.77  &  6.52  &  CEMP-no \\
                               &                           &         &           &           &        &        &          \\
HE~1243-2408\tablenotemark{a}  &  12 45 54.1 $-$24 24 46   & 10.85   &  0.81     &  $-2.84  $&  0.75  &  6.34  &  CEMP-no \\
HE~1300-2739                   &  13 03 19.8 $-$27 55 54   & 10.19   &  0.72     &  $-1.63  $&  0.89  &  7.69  &  CEMP-$s$\\
HE~1327-2116\tablenotemark{a}  &  13 30 19.4 $-$21 32 03   & 11.59   &  1.11     &  $-3.48  $&  2.64  &  7.59  &  CEMP-$s$\\
HE~1350-2955                   &  13 53 05.7 $-$30 10 11   & 10.36   &  0.49     &  $-1.14  $&  0.83  &  8.12  &  CEMP-$s$\\
HE~1403-2207                   &  14 06 41.5 $-$22 21 23   &  9.74   &  0.22     &  $-1.16  $&  0.78  &  8.05  &  CEMP-$s$\\
HE~1410-0125                   &  14 13 24.8 $-$01 39 53   & 12.61   &  1.25     &  $-2.87  $&  1.47  &  7.03  &  CEMP-no \\
HE~1412-0847                   &  14 14 57.4 $-$09 01 45   & 12.58   &  0.60     &  $-1.81  $&  1.96  &  8.58  &  CEMP-$s$\\
HE~1457-1215\tablenotemark{a,b}&  15 00 30.9 $-$12 26 57   & 10.18   &  0.55     &  $-1.56  $&  1.23  &  8.09  &  CEMP-$s$\\
BM-218                         &  15 47 47.9 $-$57 48 30   &  8.93   &  0.65     &  $-1.43  $&  0.77  &  7.77  &  CEMP-$s$\\
BM-285\tablenotemark{a}        &  21 06 02.9 $-$61 33 45   &  9.81   &  0.73     &  $-2.12  $&  0.84  &  7.15  &  CEMP-$s$\\
                               &                           &         &           &           &        &        &          \\
BM-287                         &  21 09 04.6 $-$55 17 36   &  8.35   &  0.34     &  $-1.17  $&  1.24  &  8.50  &  CEMP-$s$\\
HE~2138-0314\tablenotemark{a}  &  21 40 41.6 $-$03 01 17   & 13.23   &  0.57     &  $-3.07  $&  0.91  &  6.27  &  CEMP-no \\
HE~2155-2043\tablenotemark{a}  &  21 58 42.3 $-$20 29 16   & 13.19   &  0.75     &  $-3.27  $&  0.81  &  5.97  &  CEMP-no \\
HE~2214-1654\tablenotemark{a}  &  22 17 01.7 $-$16 39 27   & 13.19   &  0.81     &  $-3.60  $&  1.08  &  5.91  &  CEMP-no\tablenotemark{c}\\
BM-309                         &  22 37 51.0 $-$60 05 41   &  8.69   &  0.43     &  $-1.00  $&  0.74  &  8.17  &  CEMP-$s$\\
HE~2235-5058\tablenotemark{a}  &  22 38 08.0 $-$50 42 41   & 12.92   &  0.92     &  $-3.81  $&  3.16  &  7.78  &  CEMP-$s$\\
HE~2240-1647                   &  22 42 57.0 $-$16 31 20   & 12.68   &  0.78     &  $-3.18  $&  1.51  &  6.75  &  CEMP-no \\
HE~2250-4229\tablenotemark{a}  &  22 53 39.7 $-$42 13 04   & 11.91   &  0.75     &  $-2.83  $&  0.82  &  6.42  &  CEMP-no \\
HE~2319-5228\tablenotemark{a}  &  23 21 58.2 $-$52 11 43   & 13.25   &  0.90     &  $-3.39  $&  1.47  &  6.51  &  CEMP-no \\
HE~2342-3815                   &  23 45 08.3 $-$37 59 15   & 11.10   &  0.36     &  $-1.12  $&  0.83  &  8.14  &  CEMP-$s$\\
                               &                           &         &           &           &        &        &          \\
HE~2343-1817                   &  23 46 14.7 $-$18 00 47   & 11.90   &  0.64     &  $-2.14  $&  2.09  &  8.38  &  CEMP-$s$\\
\enddata

\tablenotetext{a}{A high-resolution spectrum exists for this star.}
\tablenotetext{b}{This star is also BM-209.}
\tablenotetext{b}{This star is a rediscovery of CS~22892-052, a known CEMP-$r$ star.}

\end{deluxetable}

\vfill
\newpage
\end{landscape}
\end{document}